\documentclass[10pt,journal,compsoc]{IEEEtran}
\ifCLASSOPTIONcompsoc
  \usepackage[nocompress]{cite}
\else
  \usepackage{cite}
\fi
\ifCLASSINFOpdf
\else
\fi
\usepackage{url}
\usepackage{xcolor}
\usepackage{graphicx}
\usepackage{subcaption}
\usepackage[version=4]{mhchem} %used for chemical formulae
\usepackage{booktabs} %used for table
\usepackage{makecell} %To keep spacing of text in tables
\usepackage{soul}
\usepackage{hyperref}
\usepackage[ruled,vlined,linesnumbered]{algorithm2e}
\usepackage[super]{nth}
\usepackage{amsmath}    %split math equation to fit on two lines
\usepackage{booktabs} %used for table
\usepackage{makecell} %To keep spacing of text in tables
\usepackage{multirow}
\usepackage{enumitem}

\newcommand\note[1]{\textcolor{brown}{#1}}

\SetCommentSty{mycommfont}

\begin{document}
\title{Approximate MRAM: High-performance and Power-efficient Computing with MRAM Chips for Error-tolerant Applications}
%\title{Approximate MRAM: The Role of Emerging Memory Chips in Performance and Energy Enhancement}

%------------------------------------------------------------------------------------------%
\author{Farah Ferdaus,~\IEEEmembership{Student Member,~IEEE,}
        B. M. S. Bahar Talukder,~\IEEEmembership{Student Member,~IEEE,}
        and Md Tauhidur Rahman,~\IEEEmembership{Member,~IEEE}% <-this % stops a space
%\IEEEcompsocitemizethanks{\IEEEcompsocthanksitem 
\thanks{The authors are with the Department of Electrical and Computer Engineering, Florida International University, Miami,
FL, 33174.\protect\\
E-mail: \{fferd006, bbaha007, mdtrahma\}@fiu.edu.}
%\IEEEcompsocthanksitem J. Doe and J. Doe are with Anonymous University.}% <-this % stops an unwanted space
\thanks{Manuscript received April 19, 2005; revised August 26, 2015.}}
%------------------------------------------------------------------------------------------%


\markboth{Journal of \LaTeX\ Class Files,~Vol.~14, No.~8, August~2015}%
{Ferdaus \MakeLowercase{\textit{et al.}}:
Approximate MRAM: High-efficient and Energy Efficient Computing}%Approximate MRAM: The Role of Emerging Memory Chips in Performance and Energy Enhancement}



\IEEEtitleabstractindextext{%
%------------------------------------------------------------------------------------------%
\begin{abstract}
%------------------------------------------------------------------------------------------%

Approximate computing (AC) leverages the inherent error resilience and is used in many big-data applications from various domains such as multimedia, computer vision, signal processing, and machine learning to improve systems performance and power consumption. Like many other approximate circuits and algorithms, the memory subsystem can also be used to enhance performance and save power significantly. This paper proposes an efficient and effective systematic methodology to construct an approximate non-volatile magneto-resistive RAM (MRAM) framework using consumer-off-the-shelf (COTS) MRAM chips. In the proposed scheme, an extensive experimental characterization of memory errors is performed by manipulating the write latency of MRAM chips which exploits the inherent (intrinsic/extrinsic process variation) stochastic switching behavior of magnetic tunnel junctions (MTJs). The experimental results, involving error-resilient image compression and machine learning applications, reveal that the proposed AC framework provides a significant performance improvement and demonstrates a reduction in MRAM write energy of ${\sim}47.5\%$ on average with negligible or no loss in output quality.
\end{abstract}
% \vspace{-0.3cm}
\begin{IEEEkeywords}
%------------------------------------------------------------------------------------------%
Approximate Computing, MRAM, Cache Write Policy, Low Power Design, Approximate MRAM.%, journal, \LaTeX, paper, template.
\end{IEEEkeywords}}

\maketitle

\IEEEdisplaynontitleabstractindextext

% \fi
%\vspace{-0.3cm}
\IEEEpeerreviewmaketitle

%------------------------------------------------------------------------------------------%
\IEEEraisesectionheading{\section{Introduction}\label{sec:introduction}}

% for IEEE Computer Society journal papers produced under \LaTeX\ using
% IEEEtran.cls version 1.8b and later.
% % You must have at least 2 lines in the paragraph with the drop letter
% % (should never be an issue)
% I wish you the best of success.

 
\IEEEPARstart{T}{he} amount of unstructured data processing from numerous modern computing platforms, from mobile to the cloud, is increasing rapidly. Therefore, we need an efficient framework to mitigate the gap between processor and memory systems and accomplish the computational and storage demands of the current computing system, even in advanced multi-core processors. In addition, memories consume a significant power and chip area, becoming the computing system's main bottleneck. Moreover, continual scaling down in technology nodes introduces additional challenges to the conventional memory chips, such as increased leakage power, substantial process variation, high sensitivity with different operating conditions, etc. Existing mainstream volatile memory chips, i.e., static RAM (SRAM) and dynamic RAM (DRAM), suffer from density, scalability, memory persistency, and leakage issues. Besides, current non-volatile memory (NVM) chips (e.g., Flash) suffer from endurance and performance problems. These limitations make existing memory chips incompetent for delivering ever-increasing demands of power-efficient, high-performance systems with the growing number of cores and data volume \cite{DRAM_limit}. 

Fortunately, data-intensive emerging applications in graphics/multimedia, data-driven inference, computer vision, data mining, and machine learning exhibit intrinsic error resilience properties. Therefore, those applications can produce outputs with acceptable quality in the presence of data or computation approximations \cite{AC_survey}. %\st{By leveraging such applications' intrinsic error tolerance nature,}
Approximate computing (AC) attains performance and power improvement by orders of magnitude using software, architecture, and circuit-level techniques \cite{AC_Han, AC_survey}. Therefore, %\st{these intrinsic characteristics of applications endow producing good-enough quality results even in noisy or slightly erroneous imprecise inputs. Hence,}
AC is attracting significant interest in implementing approximate memories by relaxing the strict constraints on data integrity in a controlled manner, and as a result, saves significant power consumption as well as improves performance notably.  

MRAM chip has significant potential to turn into a dominant universal memory (cache or main memory) technology due to its attractive characteristics, such as non-volatility, scalability, high speed and fast access latencies, large endurance, ultralow-power operation, CMOS compatibility, high density, reliability, near-zero leakage, almost zero static power, thermal robustness, radiation hardness, etc. \cite{Survey_Ghosh}. These advantageous features make MRAM suitable to integrate into most of the systems. Therefore, MRAM can be an excellent candidate for energy-efficient on-chip memory. However, the higher write current for a sufficiently long duration is required for proper magnetic switching during the write operation, limiting the overall energy efficiency. Several promising research has been performed on the circuit, device, and architecture levels to improve the energy efficiency of MRAM \cite{STT_Roy, STT_Jog, Hybrid_cache, AC_Tahoori}. This paper proposes a new dimension of MRAM-based approximate approach to attain significant performance and power gain by introducing a small number of errors that most error-resilient applications can endure.

The major challenge in AC is managing the approximations to trade-off between the application quality and device performance (i.e., energy, speed, etc.). Most AC-based prior works are focused on processing or logic circuits and can be categorized based on the targeted memory hierarchy level \cite{Flikker, AC_dram, AC_SRAM_Roy, AC_STT_Roy, AC_SRAM, AC_Load}. These works focus on (i) secondary storage and main memory \cite{Flikker, AC_dram}, (ii) application-specific memory designs \cite{AC_SRAM_Roy, AC_STT_Roy}, and (iii) CMOS memory-based approximate cache architecture \cite{AC_SRAM, AC_Load}. However, the existing MRAM-based AC framework is mostly simulation-based requires extension in either new instruction set architecture (ISA) or enhancement in cache replacement policies along with device-circuit-architecture remodeling \cite{AC_STT_Roy, AC_Tahoori}. Hence, these MRAM-based AC frameworks can not be easily integrated into the existing computing system due to the above-mentioned strict requirements. The previous contributions inspirit the need for real memory implementation and build the foundation of the proposed MRAM-based AC framework using commercial-off-the-shelf (COTS) MRAM chips, which require minimal or no additional hardware, is robust against operating conditions. 
%\st{Towards this end, our proposed system-level MRAM-based experimental AC framework yields significant power benefits for small probabilities of write failures by exploiting write latency variations of COTS MRAM chips.}
To this end, we propose a system-level MRAM-based experimental AC framework that works at the reduced write latency of COTS MRAM chips to improve computation speed significantly and energy efficiency with the cost of small write accuracy.

This paper presents a new approach to constructing an approximate MRAM framework. We also introduce a data allocation scheme based on the cell characterization algorithm for the proposed framework. The core idea is that at sub-optimal write latency, MRAM addresses can be split into accurate and approximate addresses through extensive error characterization of COTS toggle MRAM chips. The correlation between the obtained errors and the application-level output quality guides the proper allocation of application data to MRAM addresses. The more straightforward implementation of the proposed mechanism results in a better speed-power-quality trade-off. In summary, the key contributions of this work are as follows.

%\vspace{-0.2cm}
\begin{itemize}[leftmargin=*, topsep=0pt,itemsep=-1ex,partopsep=1ex,parsep=1ex]
  \item {We reduce the write enable ($\overline{W}$) time from the manufacturer's recommended value during the write operation to introduce errors. We extensively characterize these errors using COTS Everspin toggle MRAM chips \cite{everspin} to improve write power efficiency by exploiting the applications' error-resilient nature. The characterization results are used to derive key insights about memory errors — for example, toggling from ``1" $\rightarrow$ ``0" and ``0" $\rightarrow$ ``1" are mutually exclusive.}
  \item {Insights obtained from the characterization guide us in constructing a systematic data allocation scheme based on the application requirements to store critical and approximate data considering the occurrence and properties of MRAM errors by writing different intuitive and non-intuitive input data patterns. Subsequently, approximate data can be stored in the approximate addresses, whereas critical data must be allocated in entirely accurate memory addresses.} 
  \item {We analyze a detailed trade-off between the application-level output accuracy and system-level performance/power gains by determining optimal write latency.}
  \item {We implement our AC framework using our custom memory controller implemented on Xilinx Artix 7 (XC7A35T-1C) FPGA to manipulate different timing latency of a couple of emerging memories \cite{Alchitry}. Our experimental results show a significant improvement in the speed-power-quality trade-off, an average write speed-up of ${\sim}29\%$ and power savings of ${\sim}47.5\%$ with minimal or no loss in application quality.}
\end{itemize}
%\vspace{-0.2cm}

The rest of the paper is organized as follows. Sect. \ref{sec:rel_work} provides an overview of related prior work. Sect. \ref{sec:back} briefly overviews the organization and operating principle of MRAM chips and related preliminaries of cache writing policies. Sect. \ref{sec:method} presents the proposed AC framework, including the cell characterization algorithm. Sect. \ref{sec:result} explains the experimental setup and exhibits obtained results. Sect. \ref{sec:discuss} discusses the characterization and memory overhead required to store the erroneous memory addresses along with the proposed scheme's applicability to next-generation MRAMs. Finally, Sect. \ref{sec:end} concludes the article.

\vspace{-0.3cm}
%------------------------------------------------------------------------------------------%
\section{Related Work} \label{sec:rel_work}
%------------------------------------------------------------------------------------------%

This work correlates two distinct research areas \textendash~ emerging MRAM memories with approximate computing. Previous simulation-based MRAM-related work focuses on addressing spintronic memories' high write energy while preserving accurate read/write operations to achieve energy-efficient on-chip memories using the circuit, architectural, and device-level techniques \cite{STT_Kim, STT_Chatterjee, hybrid_Wu, hybrid_wang, STT_zhou, STT_cache_Roy, STT_Jog}. In \cite{STT_Kim}, using the bit-line voltage clamping technique, MRAM write-current asymmetry is mitigated at the circuit level to achieve the goal. Besides, in \cite{STT_Chatterjee}, a co-design methodology focusing on the bit-cell access transistor and the supply voltage is proposed to improve energy efficiency. Moreover, the MRAM cell's actual switching time detection technique is proposed to cease the unnecessary current flow immediately after a complete write \cite{STT_zhou}. Significant write energy reduction is achieved for such techniques; however, the overall write latency remains the same. On the other hand, at the architectural level, in \cite{hybrid_Wu,hybrid_wang}, the proposed hybrid CMOS-spintronic cache can selectively direct the write-intensive memory blocks to the CMOS portion while keeping the remaining blocks in the spintronic part to address the write-inefficiency of the spintronic memories. Besides, redundant memory writes are eliminated by comparing the previously stored data before the write operation \cite{STT_zhou} or tracking dirty data at a finer granularity \cite{STT_cache_Roy}. In addition, in \cite{stt_emre}, significant write margin reduction is achieved by exploiting incomplete write operations, and the incomplete bits are processed through robust Error Correction Codes (ECCs) at the cost of large decoding latencies (impairing the memory read latency). Furthermore, at the device level, in \cite{STT_Jog, stt_intel}, volatile spintronic memory is proposed by relaxing the non-volatility property (data-retention-time) or the current to overcome high write latency and energy issues of STT-MRAMs %\st{alongside the proposed refresh techniques can avoid data retention errors}
at the cost of higher write errors. However, those proposed techniques require an additional data refresh scheme to avoid further retention errors. Besides, in \cite{stt_smith}, the thermal stability factor \cite{tsf}, which defines the stability of the free layer's magnetic orientation against thermal noise, is reduced to address extreme write margin at the cost of a higher overall failure rate and retention time. 

In addition to the above techniques, a few previous AC-based frameworks explore different layers of abstraction spanning software or circuits, architectures, and algorithms to reduce the consumed energy in processing cores/accelerators. Introducing errors in the memory subsystem is an effective way of exploiting the applications' intrinsic error resilience. However, we concentrate on prior works related to approximate memories since that aligns with this work. Prior works on approximate DRAMs \cite{Flikker, AC_dram} relax their refresh rate to inject retention errors in the saved or read data contents in exchange for refresh power savings. On the other hand, \cite{AC_PCM} proposes approximate multilevel PCMs that can (i) reduce the number of programming pulses to inject write errors in exchange for energy savings and (ii) perform partial error correction on the most significant bits of the worn-out blocks. However, spintronic memories require exploring new technique due to the in-applicability of the mentioned approximation mechanisms. Moreover, \cite{AC_STT_Roy, AC_STT_Roy_DATE} proposed simulation-based enhanced architectural and software approximate design models by accessing cells in the memory arrays with different current levels depending on application requirements at the cost of higher read/write error rates. The proposed technique can (i) specify acceptable error probabilities for groups of bits within the word by controlling the quality of the read/write operations (ii) regulate (by software) the numerical significance of errors incurred during approximate load/store operations. However, this work entirely ignores the process variation effect. The process variation in spintronic memories affects both the MTJ cell and CMOS transistor parameters, which, in turn, impacts the parameters associated with read/write operations such as switching current, thermal stability factor, transistor's output current, etc. Consequently, the output accuracy at the architecture level will vary considerably with process variation due to these different and lower read/write access current levels. In \cite{AC_Tahoori}, another simulation-based new approximate spintronic on-chip memory design is proposed by relaxing both device and circuit parameters of spintronic memories by increasing various failure rates such as read disturb, retention and read decision failures, along with write errors to improve energy consumption and performance.

In contrast, our work renders a speed-power-quality trade-off through an actual memory implementation by employing COTS MRAM chips that (i) require minimal or no additional hardware, (ii) are robust against diverse operating conditions, and (iii) achieve power and performance improvements over and beyond the previous techniques, which is the key uniqueness of the proposed work.

\vspace{-0.3cm}
%------------------------------------------------------------------------------------------%
\section{Background} \label{sec:back}
%------------------------------------------------------------------------------------------%

%------------------------------------------------------------------------------------------%
\subsection{MRAM: Preliminaries}
%------------------------------------------------------------------------------------------%

\begin{figure*}[ht!]
    \centering
    \captionsetup{justification=centering, margin= 0.5cm}
    \begin{subfigure}[t]{0.35\textwidth}
        \centering
        \includegraphics[trim=0cm 10.5cm 20cm 0cm, clip, width=0.9\textwidth]{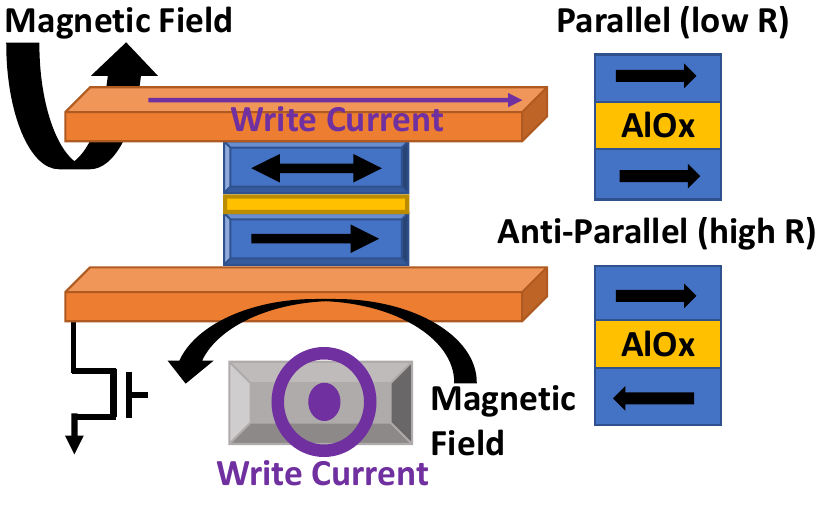}
        \caption{}
        \label{fig:MRAMcell}
    \end{subfigure}%
    \begin{subfigure}[t]{0.35\textwidth}
        \centering
        \includegraphics[trim=0cm 7.5cm 14cm 0cm, clip, width = 0.9\textwidth]{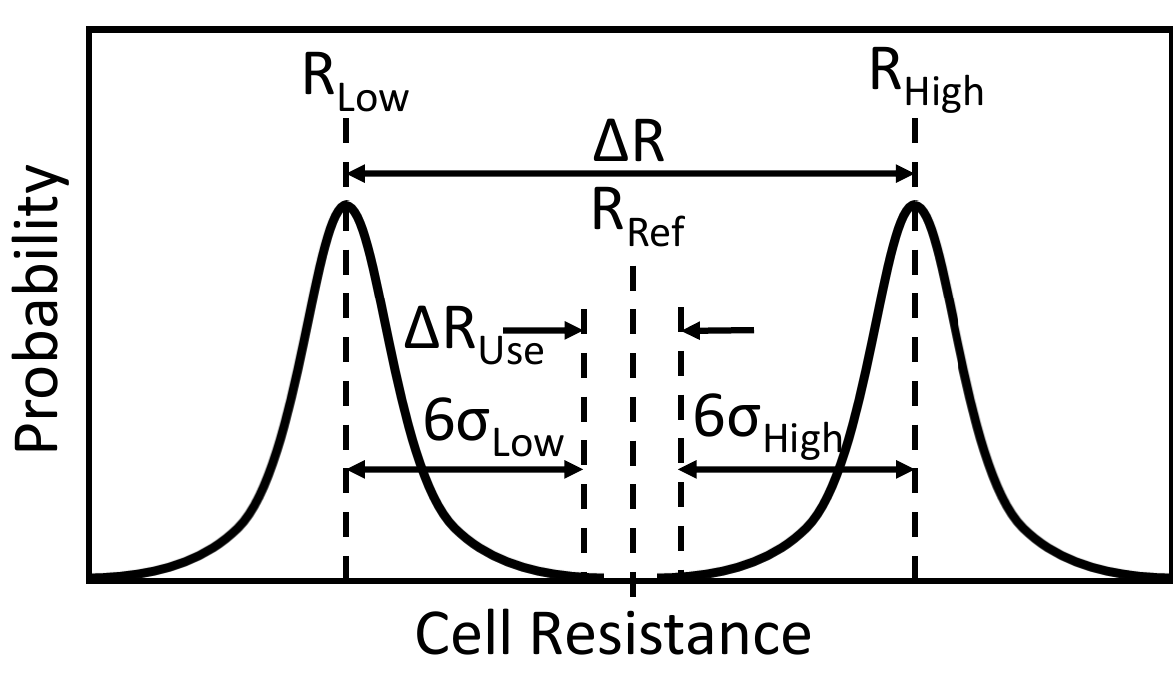}
        \caption{}
        \label{fig:cell_res}
    \end{subfigure}%
    \begin{subfigure}[t]{0.26\textwidth}
        \centering
        \includegraphics[trim=0cm 10cm 18cm 3.5cm, clip, width = 0.9\textwidth]{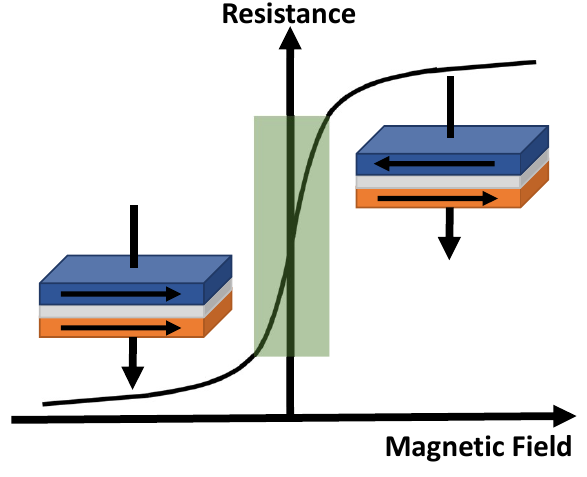}
        \caption{}
        \label{fig:TMR}
    \end{subfigure}
    %\vspace{-0.3cm}  
    \caption{(a) Toggle MRAM cell structure with MTJ. (b) Schematic representation of Gaussian resistance ($R_{Low}$ and $R_{High}$ states) distribution of larger-sized MTJ array \cite{toggle_mram}. (c) Principle of tunneling magnetoresistance (TMR) \cite{TMR}.}
    \label{fig:MRAM}
\vspace{-0.3cm}    
\end{figure*}

The core element of toggle MRAM is the magnetic tunnel junction (MTJ) that uses the Savtchenko switching \cite{toggle_mram} property %\st{by creating a rotating field with the sequential identical write current pulses}
to store both data states (high and low). The 1T-1MTJ MRAM bit cell comprises two ferromagnetic layers separated by a thin dielectric tunnel oxide ($\ce{AlOx}$ or $\ce{MgO}$) layer %\st{, built using advanced thin-film processing technology} 
(Fig. \ref{fig:MRAMcell}). One layer’s magnetic orientation is permanently fixed, commonly referred to as the reference (or fixed) magnetic layer (RML). In contrast, the other layer’s magnetization can freely be oriented depending on the magnetic field, known as the free magnetic layer (FML). %\st{The FML is composed of $\ce{NiFe}$ synthetic antiferromagnet (SAF).} 
The substantially higher magnetic anisotropy of RML compared to FML ascertains the stable magnetization direction of FML during memory read/write operation. The resistance states determine the bit that will be stored in the memory array. When both the FML and RML are aligned in the same direction% \st{(a large current passed from SelectLine (SL) to BitLine (BL) through the barrier layer)}
, the MTJ produces low electrical resistance. On the other hand, when their magnetic field orientation is opposite, the resistance becomes extremely large; hence, almost no current or weak current flows through the barrier layer. Therefore,  the MTJ exhibits high electrical resistance.

Writing a bit in the magnetic field-driven toggle MRAM array requires passing a high write current ($I_w$) for changing FML's magnetic orientation \cite{toggle_mram}. The applied $I_w$ to the write lines, placed on top and bottom of the MTJ devices (see Fig. \ref{fig:MRAMcell}), creates an auxiliary magnetic field that changes the FML direction in the required position. Contrastingly, the RML's direction is strongly coupled with an anti-ferromagnet \cite{toggle_mram}. During the write operation, the memory circuit performs a pre-read operation to determine the state of the target bit and execute a toggle pulse (if required) to change the state of the bit if the desired state is not the same as the targeted state. Consequently, it reduces the overall power consumption and improves power efficiency. However, this increases the total write cycle time (including an additional read operation).

A small bias voltage (far below the device's breakdown voltage) is applied across the MRAM cell during the read cycle. Depending on parallel ($R_{Low}$) or anti-parallel ($R_{High}$) magnetic orientation, a current sensing circuitry that is attached with the MRAM cell experiences different amounts of current and latches the appropriate logic (``0" or ``1") compared with the reference resistance ($R_{Ref}$) shown in Fig. \ref{fig:cell_res}. %\st{The random resistance variation effect of a larger-sized MRAM array read circuitry is illustrated in Fig.} \ref{fig:cell_res}. \st{The accepted bits are those whose statistical separation is greater than $5\sigma$ from the mean, where $\sigma$ is the standard deviation. The accuracy of the read circuitry entirely depends on determining the actual resistance state in the tail region (useable resistance change, $\Delta R_{Use}$) of the distribution. For robust operation, less noise-sensitive, and high-speed read operation with normal process variation, large $\Delta R_{Use}$, and significantly more than $12\sigma$ separation are crucial }\cite{toggle_mram}. \st{Furthermore,} \textcolor{blue}
{Note that} the width of resistance distribution varies from cell to cell because of manufacturing process variations. Besides, the quality, size, and level of in-homogeneity of the MTJ tunnel barrier significantly impact larger relative bit-to-bit resistance variation \cite{toggle_mram}. %\st{Therefore, a thicker tunnel barrier ($\sim1nm$) is essential to maintain the resistance level of the MTJ in the kilo $\Omega$ range for minimizing the series resistance effect from the isolation transistor }\cite{toggle_mram}, \st{where $\Omega$ is the SI unit of resistance.}

\begin{figure}[ht!]
%\vspace{-0.3cm}
    \centering
    \captionsetup{justification=centering, margin= 0cm}
    \includegraphics[trim=0cm 11.5cm 10cm 0.5cm, clip, width = 0.4\textwidth]{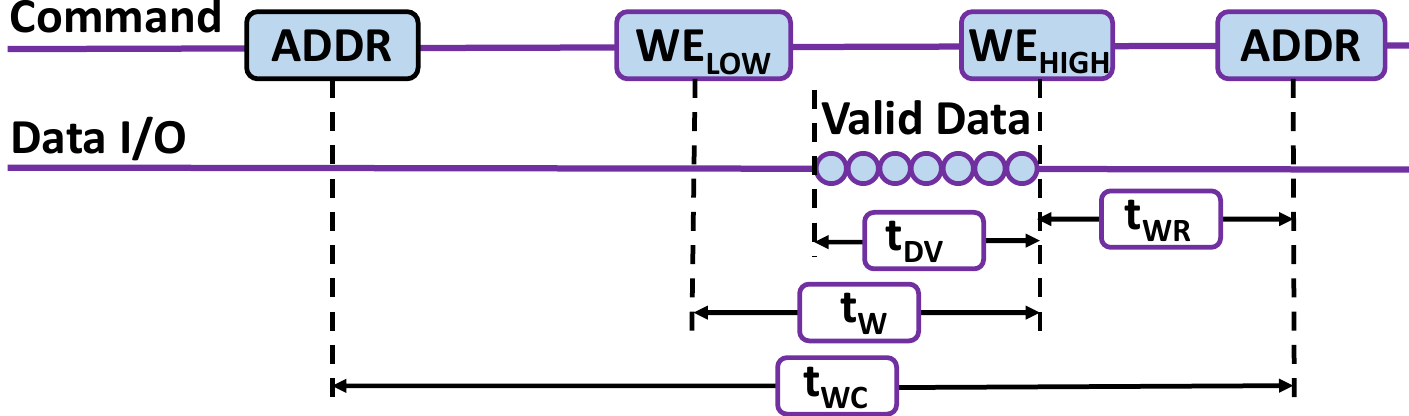}
    \vspace{-0.2cm}
    \caption{Write enable ($\overline{W}$) controlled write cycle of MRAM.}
    \label{fig:Everspin}
\vspace{-0.3cm}
\end{figure}

%\st{Magnetic state-based data storage has several advantages over charge-based storage, i.e., non-destructive read operation, no leakage during magnetic polarization, unlimited read/write endurance, no wear-out due to no movement of electrons/atoms during the switching process of magnetic polarization, etc}\cite{toggle_mram2}. \st{Besides, Savtchenko switching-based MRAM arrays possess several important performance characteristics, such as lower write error rate and fast read/write cycle ($35 ns$). Less sensitivity to external fields makes them less sensitive to manufacturing process variations} \cite{toggle_engel}. 
However, from Fig. \ref{fig:TMR}, 
% \st{we observe that the change of the induced magnetic field obtained from the high write current is steep; hence, a slight change of the induced magnetic field in the intersecting region of the resistance states can alter the decision of the read circuitry.}
we observe that the change in resistance due to the change in the induced magnetic field is steep in a certain region (the green region on Fig. \ref{fig:TMR}). In this region, a slight change in the induced magnetic field may cause a drastic change in resistance states and alter the decision of the read circuitry. Therefore, manufacturers define timing parameters for all commercial memory chips for reliable write/read operation against a wide range of operating conditions. For toggle MRAM, %\st{three different control parameters can govern the write operation of the MRAM chip:}
the write operation is governed by three different control signals: write enable ($\overline{W}$), chip enable ($\overline{E}$), and upper/lower byte enable ($\overline{UB}$/$\overline{LB}$) signals \cite{everspin}. A simplified version of the write enable ($\overline{W}$) controlled write operation of the MRAM chip is shown in Fig. \ref{fig:Everspin}.     

Here,

$t_{WC}$ = \textit{write cycle time}, i.e., the time period to complete full write operation in a particular address.

$t_{W}$ = \textit{write~pulse~width}, i.e., the time period for which the $\overline{W}$ pin is kept activated. 

$t_{WR}$ = \textit{write~recovery~time}, i.e., the time to complete the write operation after the $\overline{W}$ pin is deactivated. 

$t_{DV}$ = \textit{valid~data~to~end~of write}, i.e., the time for which the valid data need to be available in the data I/O before the $\overline{W}$ pin is deactivated.  

%\st{If the output enable ($\overline{G}$) becomes active simultaneously or after $\overline{W}$ is activated, the output will remain in the high impedance state. After all three write control parameters ($\overline{E}$, $\overline{W}$, or $\overline{UB}$/$\overline{LB}$) become disabled, the $\overline{G}$ signal must remain in the steady-state high for at least $2 ns$.} 
Reducing any of these timing parameters can improve the speed and reduce power consumption but may lead to faulty operation. The write timing parameter $t_{W}$ is manipulated in this work to introduce errors during the $\overline{W}$-controlled write operation.

%\IEEEpubidadjcol

% \subsubsection{Subsubsection Heading Here}
% Subsubsection text here.

\vspace{-0.3cm}
%------------------------------------------------------------------------------------------%
\subsection{Read vs. Write Latency}
%------------------------------------------------------------------------------------------%

%\vspace{-0.05cm}
%MRAM has the potential to be used both as the main memory and the cache. 
Due to the technology limitation, the MRAM is still much slower than the most commonly used SRAM cache ($\sim35ns$ access time on MRAM vs. ${\sim}2$--$20ns$ access time on SRAM cache). Hence, we will keep our discussion limited by considering MRAM as the main memory in this section. However, it is worth mentioning that our proposed reducing \textit{MRAM write latency} strategy will also improve the system performance, even if it is used as a cache.

In a computer system, the main memory is only accessed when there is a cache miss (read/write miss). Usually, recovering a read miss is a direct procedure; i.e., when a read miss occurs, the corresponding memory block (usually consists of multiple data words) of the main memory is copied directly to the cache. However, by taking proper caution at the software level, one can maximize the cache hit rate and hide the read access latency of the main memory. For example, the average cache miss rate in a modern computer processor is $<$8\% for most use-cases \cite{Kumar_cache}. On the other hand, handling a write miss is more complex. Usually, most of the modern processors use either of the following two policies- (i) write-through policy with the no-write allocation and (ii) write back policy with write allocation \cite{hennessy_Archi}. The first policy is not used in high-end processors since %\st{, when a write miss occurs, the corresponding data are directly modified both on the cache and the main memory. Although this procedure is much simpler, it}
it requires a long time to proceed. %\st{On the contrary, the second policy is a three-step procedure- (i) at first, the corresponding memory block is copied from the main memory to the cache, (ii) then, the processor only updates the target word in the newly copied cache block, (iii) lastly, when the cache is required to be evicted, the corresponding memory block is written back to the main memory.} 
However, both cache miss policies become more complex if some of the memory resources are shared by the multiple processor cores (e.g., L2/L3 caches and the main memory) and require a dedicated cache coherency protocol. As the write miss recovery procedure is more complex than the read miss, the write miss penalty is much higher than the read miss penalty. Moreover, the write miss penalty might become a critical bottleneck for write-intensive program performance (e.g., iterative weight updating on neural networks). Moreover, write operations consume a significant portion of total energy\cite{AC_Tahoori, AC_STT_Roy}. Hence, in this work, we aim to reduce such penalties by improving the write access time of MRAMs.

Furthermore, big data applications allow data approximation to some extent, motivating us to develop an AC framework to investigate the trade-off between application-level acceptable output accuracy vs. system-level power/performance gain in the memory system.
%\vspace{-0.3cm}

%------------------------------------------------------------------------------------------%
\vspace{-0.3cm}
\section{MRAM Approximate Computation Framework} \label{sec:method}
%------------------------------------------------------------------------------------------%

Random Savtchenko switching property is exploited at the reduced timing parameter in our proposed methodology to obtain data approximation at the application level. At reduced \textit{write pulse width}, $t_{W}$, of toggle MRAM (see Fig. \ref{fig:Everspin}),  all memory cells do not receive sufficient write current and time to toggle into the intended stable state. These random variations are created in the MTJ storage element due to the process variation and the non-uniform distribution of the current pulse within the chip, which hinders performing an appropriate write operation in all memory cells. That is why the manufacturer specifies a set of timing parameters for reliable read/write operations. Hence, violation in any of these manufacturer-recommended timing parameters may cause erroneous/faulty outputs during the read/write operation. If the $t_{W}$ is insufficient, there is a high chance that FML cannot align perfectly with the RML (either the same or opposite direction) and might be settled on an intermediate position, leading to the cell resistance being halfway between $R_{Low}$ and $R_{High}$ \cite{TMR}. Therefore, at reduced $t_{W}$, if the resultant cell resistance falls around the $\Delta R_{Use}$ region of the resistance distribution (see Fig. \ref{fig:cell_res}) curve, the cell will show indeterministic characteristics and generate erroneous bits.

Several steps are involved in our proposed scheme. At the reduced $t_{W}$ , MRAM chips create errors, and the total number of errors differs at different reduced $t_{W}$ values. At first, we select the most suitable reduced $t_{W}$ value. This selected $t_{W}$ aims to generate erroneous bits at a tolerable range for error-resilient applications. Second, we propose a cell selection algorithm to characterize all MRAM cells from a set of measurements by writing a specific data pattern that produces maximum error compared with other data patterns to identify the erroneous (approximate) and error-free (accurate) memory cells/addresses for the AC framework. The characterization needs to be performed only once to choose the appropriate number of MRAM cells/addresses. Finally, we collect data by writing all necessary data patterns into the entire memory to gain valuable insights about the bit errors' frequency, significance, and nature.

%\vspace{-0.3cm}
%------------------------------------------------------------------------------------------%
\subsection{Appropriate Reduced Time Selection}
%------------------------------------------------------------------------------------------%

Appropriate reduced time selection is essential for obtaining the most favorable trade-off between application quality and speed/power improvement. Towards this end, the experimental results reveal that some of the memory cells provide erroneous outputs if the data is written at the reduced timing parameters \cite{mram_trng}. The number of these error-prone cells varies within the write pulse activation time range t = [0~ $t_{W}$]. The total number of erroneous bit cells is counted by changing the $t_{W}$ and  writing different input data patterns at different $t_{W}$. The main objective is to determine a suitable $t_{W}$ for which the acceptable amount of erroneous bits is achieved. In the next step, the number of erroneous cells is calculated from all achievable reduced write timing parameters. Finally, we propose an algorithm to characterize the erroneous memory cells for the AC framework using the timing parameter for which the sustainable amount of error bits is obtained.

\vspace{-0.3cm}
%------------------------------------------------------------------------------------------%
\subsection{Erroneous Address Selection} \label{subsec:cell_char}
% \subsection{Erroneous Address Selection through MRAM Cells Characterization} \label{subsec:cell_char}
%------------------------------------------------------------------------------------------%

Our experimental result manifests that only a few memory cells, hence addresses are error-prone at a specific sub-optimal write latency. To locate these cells, we characterize MRAM memory cells by writing different intuitive (solid) and nonintuitive (random, checkerboard, and striped) input data patterns to the entire memory cells at the reduced write enable time, $t_{W}$, and read back the entire memory contents with appropriate timing parameters a total of $N$ times. Larger $N$ provides better characterization results, but it increases the computation time.

%\st{Theoretically, reduced write operation reduces the current flowing through the MTJ storage elements; hence, the magnetic direction (\textit{parallel (P)} $\rightarrow$ \textit{anti-parallel (AP)} or vice versa) toggling time increases significantly} \cite{redu_write}. 
To toggle magnetic direction, \textit{parallel (P)} $\rightarrow$ \textit{anti-parallel (AP)}, or vice versa, in the MRAM cell, the write current ($I_{w}$) needs to be held sufficiently long \cite{redu_write}. If we reduce the write time (i.e., $\overline{W}$ pulse width), the cell resistance might be stuck in an intermediate value (anywhere between $R_{Low}$ to $R_{High}$) and might occur data corruption. However, while toggling the cell resistance ($R_{Cell}$), switching the side of the $R_{Ref}$ (i.e., $R_{Low} \rightarrow R_{Ref} + \delta$ or  $R_{High} \rightarrow R_{Ref} - \delta$) is sufficient to write the data appropriately (see. Fig. \ref{fig:cell_res}). Although just switching side of $R_{Ref}$ might affect the long-term retention capability, that should not affect the functionality of MRAM as a cache/main memory as long as the same memory component is not used as the storage memory. In the characterization phase, we aim to find the appropriate value of $\overline{W}$ pulse width, which toggles $R_{Cell}$ from one side of the $R_{Ref}$ to another side for the majority of MRAM cells. Note that switching from \textit{P} ($R_{Low}$) to \textit{AP} ($R_{High}$) is more vulnerable to reduced write operation due to enhanced switching delay, leading to write failures. Likewise, from preliminary MRAM characterization with different data patterns and different reduced  $t_{W}$, we observe the followings:

%\vspace{-0.2cm}
\begin{enumerate}[leftmargin=*, topsep=0pt,itemsep=-1ex,partopsep=1ex,parsep=1ex]
  \item {Error patterns and their nature and frequency depend entirely on the input data pattern to be written and vary with different memory chips.}
  \item {Experimental results also manifest that the write operation at the reduced $t_{W}$ produces comparatively more erroneous data for \textit{solid} \texttt{0x0000} data patterns.}
  \item {Silicon results further reveal that errors occur due to random variations in the MRAM having no relation to its internal hardware implementation.}
\end{enumerate}
%\vspace{-0.2cm}

Furthermore, error patterns from the sample measurements reveal that
% errors occur at random locations in arbitrary addresses resulting in unpredictable errors and equally unpredictable degradation in output application quality. Hence, 
allocating data randomly to the MRAM at reduced $t_{W}$ can degrade application-level output quality and cause complete or partial application failure, necessitating a systematic method to track the errors on different addresses and allocate the application data according to their criticality. Toward this end, we split the entire memory space into accurate and approximate addresses based on the error characteristics to efficiently allocate addresses of different applications to tolerate different error amounts for generating acceptable output quality. These accurate addresses can be coalesced logically into a required contiguous entirely accurate memory chunk by only tracking the approximate addresses for critical data allocation (devoid of any error) by noting the error characteristics.

For the proposed AC framework, addresses that contain erroneous cells need to be filtered. At first, we discover all erroneous cells, hence addresses, at the reduced $t_{W}$ from ($N$) measurements using the data pattern that produces maximum errors. %\st{Next, we count and locate the total number of bit errors (if any) in each memory address, comparing the data to be written at reduced $t_W$ and reading back the written data at the appropriate timing parameter, and categorize them (lines 1 - 9 of Algo. \mbox{\ref{alg:mramChar}}) accordingly taking the union of the error-prone addresses that occur across different measurements.} 
Next, we count and locate the total number of bit errors (if any) in each memory address (lines 1 - 9 of Algo. \ref{alg:mramChar}). To do so, we write fixed data (write-data) to the memory at reduced $t_W$ and read it back with the appropriate timing parameter. Then we compute the error comparing the write- and read-data. Lastly, we categorize the error by taking the union of the error-prone addresses across different measurements.
The pseudo-code for erroneous address selection and accumulation (strategy 1) and sorting (strategy 2) is shown in Algorithm \ref{alg:mramChar}. The sorting of erroneous addresses is performed based on the stored data (\textit{sD}) and intended data to be written (\textit{oD}) (lines 11 - 18). The entire AC framework methodology is split into two phases: training and evaluation. Error-prone addresses are accumulated from $N$ measurements using the most erroneous input data pattern in the training phase. In the evaluation phase, error statistics and output quality are analyzed using different input data patterns and testing big data applications.  

%\vspace{-0.3cm}
\begin{algorithm}[ht!]
\SetAlgoLined
    \KwData{
        $N$: \note{Number of total measurements.} \break 
        $A$: \note{Set of addresses in MRAM.} \break
        $\mathcal{E_A}$: \note{Set of total erroneous addresses from $N$ measurements.} \break
        $wL$: \note{Word Length} \break
        $oD$: \note{$(A \times wL)$ matrix containing data intended to write each memory cells} \break
        $sD$: \note{$(A \times wL)$ matrix containing data stored to each memory cells at reduced $t_W$} \break
        $\mathcal{E_A^Q}$: \note{$\mathcal{E_A}$ where higher value of $\mathcal{Q}$ represents lower quality level at reduced $t_W$.} \break
        $P_{i}$: \note{Pattern used to initialize memory.} \break
        $P_{t}$: \note{Pattern to test the memory.} \break
        $t_{W}^{n}$: \note{nominal value of $t_{W}$.} \break
        $t_{W}^{r}$: \note{reduced value of $t_{W}$.}}
    \BlankLine
    \KwResult{\note{Characterized MRAM}}
    \BlankLine
    \tcp{Initialization} 
    $write\_mem(P_{i}, t_{W}^{n});$ \tcp{Initialize MRAM}
    $write\_mem(P_{t}, t_{W}^{r});$ \tcp{Write MRAM with $P_{t}$ at $t_{W}^{r}$}
    $\mathcal{E_A} = \{ \}; \mathcal{E_A^Q} = \{ \};$ \break
    \hspace*{-0.5cm}\tcp{Strategy 1}
    \For{$i = 1$ to $N$}{
         \ForEach{a $\in$ A}{
         \tcp{Check write error at address $a$}
          \If{check\_error (a)}{
            \If{a $ \notin \mathcal{E_A}$}{ 
                $\mathcal{E_A} = \mathcal{E_A} \cup \{a\}$\;
                }
            }
         }
    }
    $\mathcal{E_A^Q} = sort\_addr(sD, \mathcal{E_A});$ \tcp{Only for Strategy 2}
    \BlankLine
    \SetKwFunction{FFunc}{sort\_addr}
    \SetKwProg{Fn}{Function}{:}{}
    \Fn{\FFunc{$sD, \mathcal{E_A}$}}{
        $oD = empty()$\;
        $oD = init()$; \tcp{Initialize} \break
        $x = bitwise\_xor(oD, sD)$\;
        $x = bin2Dec(x,dim=2)$\;
        $idx = argsort(x)$\;
        $\mathcal{E_A}^{sort} = \mathcal{E_A}[idx]$\;
        \KwRet $\mathcal{E_A}^{sort}$}
 \caption{Pseudo-code for accumulating erroneous addresses through  error characterization}
 \label{alg:mramChar}
%\vspace{-0.3cm}
\end{algorithm}
%\vspace{-0.3cm}
 
\begin{figure}[ht!]
%\vspace{-0.4cm}
    \centering
    \captionsetup{justification=centering, margin= 0.5cm}
    \includegraphics[trim=0cm 10cm 20.5cm 0cm, clip, width = 0.38\textwidth]{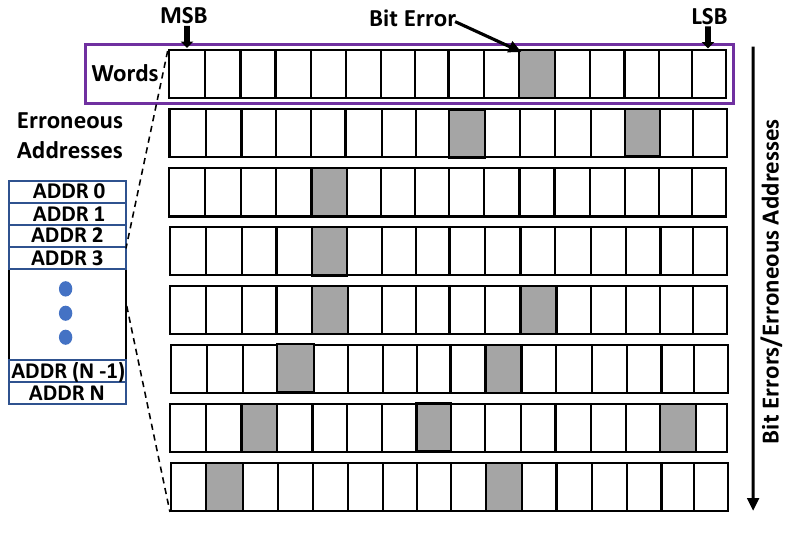}
    \vspace{-0.3cm}
    \caption{Strategy 2 for sorting erroneous addresses.}
    \label{fig:errAddrSel}
\vspace{-0.3cm}
\end{figure}

In strategy 2, we sort all the error-prone MRAM addresses in ascending order with decreasing quality (increasing number of errors) according to the total number of bit errors present in each word based on the occurrence and bit positions of the errors. A larger number of bit errors at the most significant bits (MSB) position denote higher application quality degradation; therefore, those addresses are assigned to lower-level positions. Besides, bytes/half-words can also be chosen instead of words (if required) as the error granularity. This decision depends entirely on the stored data's nature and can be trivially extended. Fig. \ref{fig:errAddrSel} depicts the sorting strategy. Data allocation is performed to each address in the sorted order, i.e., data is first assigned to the least erroneous addresses and later to addresses with higher errors ensuring that at reduced $t_W$ the stored data always incurs the least amount of errors. However, each application's critical data must be allocated to entirely accurate addresses. Experimental results reveal that the enormous amount of accurate addresses is sufficient for most applications. Next, approximate data can be allocated to either accurate (based on availability) or erroneous addresses sorted according to the number of errors. As (i) sorting the erroneous addresses require extra overhead and complexity and (ii) most of the big data applications, the performance of strategy 1 is sufficient enough (see Sect. \ref{subsec:res_C}), so we propose strategy 2 in this work for further performance enhancement\footnote{Results of strategy 2 are not presented in this work as we achieve almost $100\%$ accuracy only using strategy 1.} (if required). %\st{for any potential suitable future error-resilient applications.} 

\vspace{-0.3cm}
%------------------------------------------------------------------------------------------%
\section{Experimental Setup, Results, \& Analysis} \label{sec:result}
%------------------------------------------------------------------------------------------%

The preliminary analysis is performed over ten (2 chips from each \textit{MR0A16ACYS35, MR0A16AYS35, MR1A16AYS35 MR2A16ACYS35, MR2A16AYS35} models) 16-bit parallel interfaced differently sized ($1 Mb - 4 Mb$) toggle MRAM chips to verify our proposed scheme. These chips belong to Everspin technologies.
% Table \ref{Tab:MRAM_Char} captures the key features of the memory chips. 
Among them, five chips are selected randomly to perform extensive analysis to verify our observations' true universality. The characterization is performed using our custom memory controller implemented on the Alchitry Au development board containing \textit{Xilinx Artix 7 (XC7A35T-1C)} FPGA to manipulate different timing latency of a couple of emerging memories \cite{Alchitry}. The temperature (low/high) experiments were performed by placing the memory socket that holds the memory chip in \textit{Temptronic ATS-605 ThermoStream} to maintain a uniform temperature. %\st{Furthermore, the current analysis was performed using a $100 MHz$ \textit{Siglent CP5030A} current probe and monitored with a $1 GHz$ \textit{Siglent SDS5104X} oscilloscope.}

According to Sect. \ref{subsec:cell_char}, the generated errors are pattern-dependent on reduced write operation. Therefore, to determine the suitable pattern and examine the data pattern dependency, a total of 5-set measurement data is collected with seventeen different intuitive (solid) and non-intuitive (random, checkerboard, and striped) 16-bit input data patterns: (\texttt{0xFFFF, 0xAAAA,0x5555, 0x0000}) from each ten memory chips\footnote{The least significant 8 bits of the address-bit are used for column addressing for all memory chips.} before performing characterization. We observed that, at reduced $t_{W}$, switching from ``1" to ``0" produces more bit-error in our test MRAM chips. Consequently, the reduced $t_{W}$ produces maximum bit-error with the \textit{solid} \texttt{0x0000} pattern when the memories are initialized with logic ``1". Therefore, we initialized memory with ``1" for all test patterns (except for the \textit{solid} \texttt{0xFFFF} test pattern) to simulate the worst possible usage scenario. %However, we initialized memory with ``0" for the \texttt{0xFFFF} test pattern. 
Based on the observation and the discussion on Sect. \ref{subsec:cell_char}, we also conclude that the \textit{parallel (anti-parallel)} configuration is the logic state '1' (state '0'). Next, to characterize the MRAM cells (discussed in Sect. \ref{subsec:cell_char}), we collected a total of 50-set measurement data with only \textit{solid} \texttt{0x0000} input pattern from the randomly selected five memory chips. We chose the value of $t_W$, $33.3\%$ of the recommended $t_W$, for this work to obtain approximate writes. However, our selected value of $t_W$ can generate a sufficient number of accurate addresses (a moderate number of incorrect outputs) for the proposed framework.

%\vspace{-0.3cm}
% %
% \begin{table}[ht!]
% \caption{MRAM Chip Specifications \textcolor{green}{Can't we remove this table?} \cite{everspin}.}
% \setcellgapes{2pt}%parameter for the spacing
% \captionsetup{justification=centering, margin= 0cm}
% %\vspace{-0.3cm}
% \makegapedcells
% \centering
% \setlength\tabcolsep{2pt} 
% \resizebox{0.41\textwidth}{!}
% {
%     \begin{tabular}{|l|r|}
%     \hline
%     \multicolumn{1}{|c|}{Parameter} & \multicolumn{1}{c|}{Standard Value} \\ \hline
%     Capacity, Supply Voltage                        & 1 - 4 Mbit, 3.3 V                 \\ \hline
%     Read/Write Cycle ($t_{WC}$)     & 35 ns                      \\ \hline
%     Write Pulse Width ($t_W$)       & 15 ns                      \\ \hline
%     Write Recovery Time ($t_{WR}$)  & 12 ns                      \\ \hline
%     Valid Data to end of Write ($t_{DV}$) & 10 ns                      \\ \hline
%     Address/Data Bus Length         & 16 - 18/16                 \\ \hline
%     Retention Time                  & \textgreater 20 years      \\ \hline
%     AC stand by Current             & 18–28 mA                   \\ \hline
%     AC Active Current (Read/Write)  & 55–80 mA/105–165 mA        \\ \hline
%     \end{tabular}
% }
% \label{Tab:MRAM_Char}
% %\vspace{-0.3cm}
% \end{table}
% %

Analyzing results from the conducted experiments validate our proposed scheme's novelty and provide interesting insights. The results are divided into five broad subsections. In the first part, we manifest the speed-quality (number of erroneous cells/addresses) trade-off for selecting the reduced $t_W$ appropriately, a universal trait in any approximate computing mechanism. The second part presents the characterization results in detail applying Algo. \ref{alg:mramChar}. The third part shows how the output quality is affected when data is written into the memory with different initialized values and with or without applying the proposed address selection strategy (strategy 1 of Algo. \ref{alg:mramChar}). We also show that our approach results in comparably higher performance improvement than many popular (non) volatile memory-based prior works \cite{AC_Tahoori, Flikker, AC_dram, AC_STT_Roy}. Moreover, an efficient implementation of a memory controller can further improve the overall performance of our proposed framework. Note that all results are presented in the Sects. \ref{subsec:res_B} and \ref{subsec:res_C} are obtained at room temperature ($26^{\circ}C$). The fourth part describes the approximate MRAM results subject to operating conditions and memory chip variations. Finally, the fifth part elucidates our proposed technique's power analysis results. 

\vspace{-0.3cm}
%------------------------------------------------------------------------------------------%
\subsection{Selection of \texorpdfstring{$t_W$}{Lg}}\label{subsec:res_A}
%------------------------------------------------------------------------------------------%

\begin{figure}[ht!]
\vspace{-0.4cm}
    \centering
    \captionsetup{justification=centering, margin= 0cm}
    \includegraphics[trim=0.25cm 0.25cm 0.25cm 0.25cm, clip, width = 0.49\textwidth]{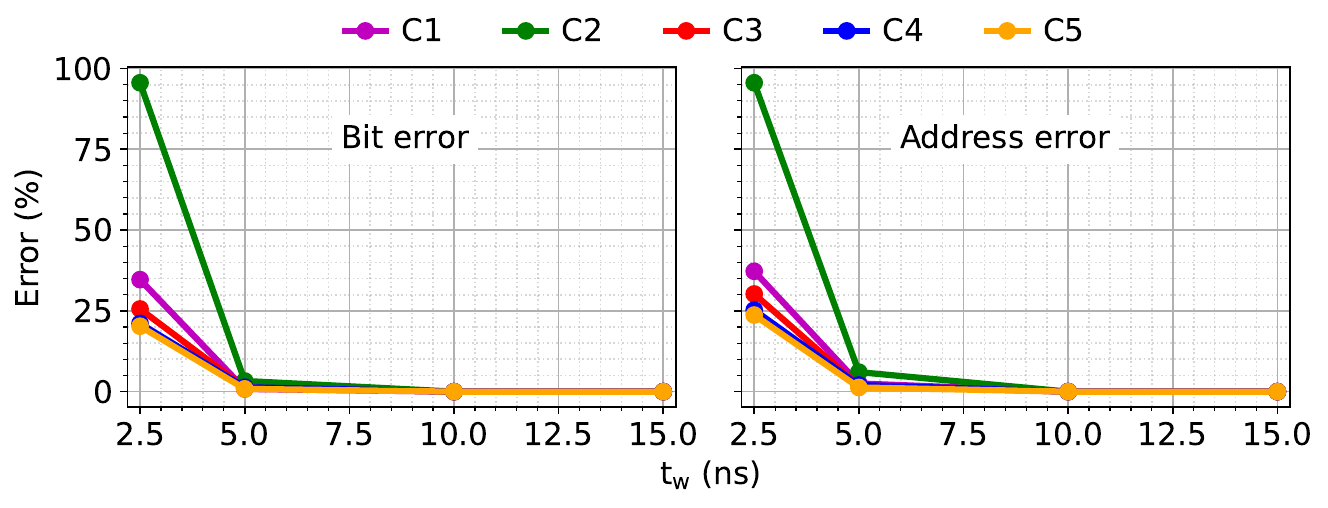}
    \vspace{-0.6cm}
    \caption{$t_{W}$ vs. errors for five randomly chosen chips. The left and right figures present \% of erroneous bits and \% of addresses which consist those erroneous bits, respectively.}
    \label{fig:select_tw}
%\vspace{-0.3cm}
\end{figure}

An extensive analysis is performed to determine the error profile, such as frequency, location of the bit errors, comparing the behavior of the faulty/erroneous outputs at different reduced $t_W$ values using \textit{solid} \texttt{0x0000} data pattern. We reduce the $t_W$ value from $15 ns$ (manufacturer's recommended) to  $10 ns$, $5 ns$, and $2.5 ns$, respectively (Fig. \ref{fig:select_tw}). %\st{Due to the experimental setup limitations, we are incapable of reducing the $t_W$ value any further.} 
Note that our experimental setup's timing resolution is limited to $2.5 ns$. At $t_W$ = $2.5 ns$, the total failed bit counts fall within ${\sim}20\%$ -- ${\sim}95\%$, which spans over ${\sim}23\%$ -- ${\sim}95\%$ addresses. Our analysis shows that 
% many 
higher bit errors can be used for random number generation and physical unclonable function \cite{mram_trng} but is not suitable for approximate computing. 
% \textcolor{green}{(data approximation or approximate computation?)}
However, at $t_W$ = $5 ns$ and $10 ns$, the obtained total failed bits are almost negligible (${<}5\%$ and ${<}1\%$, respectively) for all ten chips (considering a single measurement). Hence, we choose $t_W$ = $5 ns$ (considering the number of failed bit counts) to characterize erroneous cells to improve write speed (${\sim}29\%$ overall improvement) and power savings (see Sect. \ref{subsubsec:powImpv}). %\st{Fig. \mbox{\ref{fig:Curr_Ana}} portrays the write current observed in different $t_W$ values. It shows that a suitable decrease in $t_W$ decreases notable power consumption by maintaining a sufficient number (see Table \mbox{\ref{Tab:Char1}}) of accurate addresses (and hence output quality).}

\vspace{-0.4cm}
%------------------------------------------------------------------------------------------%
\subsection{Characterization of Error-Prone Cells}\label{subsec:res_B}
%------------------------------------------------------------------------------------------
The cell characterization of MRAM is performed according to Sect. \ref{subsec:cell_char} with $N = 50$ temporal measurements. \textit{Solid} \texttt{0x0000} is used as the write data pattern. Moreover, the entire memory is reset with the \textit{solid} \texttt{0xFFFF} data pattern before every measurement to erase previously written data traces. In Table \ref{Tab:Char1}, the first two rows show the statistics of erroneous MRAM addresses and bits after performing the proposed cell characterization algorithm (strategy 1 of Algo. \ref{alg:mramChar}). Here, $\mathcal{E_A}$ is the percentage of total erroneous addresses,  and $\mathcal{E_B}$ is the percentage of total erroneous bits accumulated from 50 temporal measurements (obtained through characterization). The results manifest that the total number of erroneous bits obtained at reduced $t_W$ ($5 ns$) varies from chip to chip. We also observe that only a few addresses hold these erroneous bits. The rest of the rows represent erroneous addresses and bits from different data patterns and to which extent they belong to $\mathcal{E_A}$  and $\mathcal{E_B}$. Here, $\mathcal{M_A}$ is the percentage of total erroneous addresses, and $\mathcal{M_B}$ is the percentage of total erroneous bits from a specific test pattern (from a single measurement), and $\mathcal{C_A}$ is the percentage of $\mathcal{M_A}$ belonging to $\mathcal{E_A}$. Similarly, $\mathcal{C_B}$ is the percentage of $\mathcal{M_B}$ belonging to $\mathcal{E_B}$.

% \textcolor{cyan}{In Table \ref{Tab:Char1},} we observe that the erroneous addresses range from only $[5.33\% ~26.41\%]$ \textcolor{blue}{(\nth{1} row of characterization segment) and the erroneous bits range from $[3.89\% ~22.06\%]$ (\nth{2} row of characterization segment) for different memory chips at $t_W = 5 ns$.
% \textcolor{green}{First and second row of what? Table or memory?}} %after performing characterization at $t_W = 5 ns$.
Table \ref{Tab:Char1} represents that, after performing characterization using $N = 50$ temporal measurements with \textit{solid} \texttt{0x0000} data pattern, the cumulative amount of erroneous addresses is $22.41\%~(\mathcal{E_A})$, and bits is $10.49\% ~(\mathcal{E_B})$ for chip 1 (C1). During the testing phase, the number of erroneous addresses observed using \textit{solid} \texttt{0x0000} data pattern is only $2.52\%~(\mathcal{M_A})$, and bits is only $0.83\%~(\mathcal{M_B})$. Now, among these $2.52\%$ erroneous addresses, $96.92\%~(\mathcal{C_A})$ of addresses belong to the $22.41\%~(\mathcal{E_A})$, and $85.48\%~(\mathcal{C_B})$ of bits belong to the $10.49\%~(\mathcal{E_B})$. 

In summary, the cumulative maximum amount of erroneous addresses is $26.41\%~ (\mathcal{E_A})$, whereas the minimum is only $5.33\%$ (\nth{1} row of characterization segment in Table \ref{Tab:Char1}) for different memory chips at $t_W = 5 ns$. Similarly, in Table \ref{Tab:Char1}, the cumulative maximum amount of erroneous bit is $22.06\%~ (\mathcal{E_B})$, whereas the minimum is only $3.89\%$ (\nth{2} row of characterization segment). This indicates that ${\sim}75\%$ 
% \textcolor{green}{(why 78\% inside the first bracket?)}
of the addresses (${\sim}78\%$ of bits) are entirely accurate in the worst case scenario, resulting in negligible or no loss in output quality. %\textcolor{cyan}{Similarly, $\sim$$78\%$ of bits are entirely accurate, even in the worst case.} 
As the percentage of error-prone addresses is fairly small ($\sim$$25\%$) regardless of memory size,  storage overhead is minimal for storing these few memory addresses' information.

For pattern \texttt{0xAAAA} or \texttt{0x5555}, we do not observe any error for any data pattern. Based on the internal architecture of MRAM, there might be different reasons behind not getting any error. For example, if all of the I/O buffers are driven by the same power rail, the inrush current caused by the I/O buffers should be lower for these specific data patterns as only 8 out of 16 bits experience the switching \cite{inrush}. Hence, the overall timing characteristics might be improved \cite{inrush}. Moreover, the maximum obtained erroneous address ($\mathcal{M_A}$) is $6.04\%$, whereas the minimum is only $0.16\%$ for any given pattern considering a single test measurement. Similarly, the maximum obtained erroneous bit ($\mathcal{M_B}$) is $3.30\%$, whereas the minimum is only $0.08\%$.
% Moreover, the obtained erroneous addresses, $\mathcal{M_A}$ (bits, $\mathcal{M_B}$) in reduced $t_W = 5ns$, are in the range $[0.16\%~6.04\%]~([0.08\%~3.30\%])$ \textcolor{blue}{for any given pattern.} %\st{for any other data patterns considering all memory chips, while most of these belong/close to the lower limit.}
Besides, our proposed characterization technique covers most erroneous addresses ($\mathcal{C_A}$) and bits ($\mathcal{C_B}$) for all data patterns. We observed that the lowest covered erroneous address ($\mathcal{C_A}$) is $1.74\%$ and bit ($\mathcal{C_B}$) is $2.76\%$ for \textit{solid} \texttt{0xFFFF} data pattern. 
% Although the lowest covered erroneous addresses ($\mathcal{C_A}$) (and bits ($\mathcal{C_B}$)) is $1.74\%~(2.76\%)$ \textcolor{green}{(what's 2.76\% and why did you use ~ symbol? Did you want to use $\sim$ ? )}for the \textit{solid} \texttt{0xFFFF} data pattern, 
However, a closer observation reveals that \textit{solid} \texttt{0xFFFF} data pattern produces only $0.26\%$ erroneous addresses and $0.23\%$ 
% \textcolor{green}{(please clarify all numbers that are inside the first bracket?)}
erroneous bits, which is negligible ($<0.5\%$). This insignificant number of errors can be overlooked for any big data application.

\begin{table}[ht!]
\caption{Error Characteristics.}
\setcellgapes{1pt}%parameter for the spacing
\captionsetup{justification=centering, margin= 0cm}
% \vspace{-0.3cm}
\makegapedcells
\centering
\setlength\tabcolsep{5pt} 
\resizebox{0.48\textwidth}{!}
{
    \begin{tabular}{|c|c|c|c|c|c|c|c|}
    \hline
    \multicolumn{2}{|c|}{Sample Chip\footnotemark}                                                                & (\%) & C1    & C2    & C3    & C4    & C5    \\ \hline
    \multicolumn{2}{|c|}{\multirow{2}{*}{Characterization}}                                          & $\mathcal{E_A}$ & 22.41 & 26.41 & 10.34 & 8.36  & 5.33  \\ \cline{3-8} 
    \multicolumn{2}{|c|}{}                                                                           & $\mathcal{E_B}$ & 10.49 & 22.06 & 7.60  & 5.15  & 3.89  \\ \hline
    \multirow{8}{*}{Solid}                                                   & \multirow{4}{*}{\texttt{FFFF}} & $\mathcal{M_A}$ & 0.26  & 0.26  & 0     & 0     & 0     \\ \cline{3-8} 
                                                                             &                       & $\mathcal{M_B}$ & 0.11  & 0.23  & 0     & 0     & 0     \\ \cline{3-8} 
                                                                             &                       & $\mathcal{C_A}$ & 1.74  & 60.81 & ----  & ----  & ----  \\ \cline{3-8} 
                                                                             &                       & $\mathcal{C_B}$ & 2.76  & 54.39 & ----  & ----  & ----  \\ \cline{2-8} 
                                                                             & \multirow{4}{*}{\texttt{0000}} & $\mathcal{M_A}$ & 2.52  & 6.04  & 1.83  & 2.04  & 1.29  \\ \cline{3-8} 
                                                                             &                       & $\mathcal{M_B}$ & 0.83  & 3.30  & 1.25  & 1.36  & 0.86  \\ \cline{3-8} 
                                                                             &                       & $\mathcal{C_A}$ & 96.92 & 99.28 & 95.26 & 86.23 & 94.34 \\ \cline{3-8} 
                                                                             &                       & $\mathcal{C_B}$ & 85.48 & 98.36 & 94.37 & 80.73 & 91.5  \\ \hline
    \multirow{8}{*}{\begin{tabular}[c]{@{}c@{}}Row\\ Striped\end{tabular}}    & \multirow{4}{*}{\texttt{FFFF}} & $\mathcal{M_A}$ & 1.99  & 3.52  & 0.72  & 0.82  & 0.62  \\ \cline{3-8} 
                                                                             &                       & $\mathcal{M_B}$ & 0.67  & 2.13  & 0.53  & 0.56  & 0.47  \\ \cline{3-8} 
                                                                             &                       & $\mathcal{C_A}$ & 97.78 & 99.55 & 95.2  & 92.87 & 95.78 \\ \cline{3-8} 
                                                                             &                       & $\mathcal{C_B}$ & 90.9  & 98.34 & 94.42 & 91.43 & 94.11 \\ \cline{2-8} 
                                                                             & \multirow{4}{*}{\texttt{0000}} & $\mathcal{M_A}$ & 0.75  & 2.63  & 0.94  & 1.09  & 0.67  \\ \cline{3-8} 
                                                                             &                       & $\mathcal{M_B}$ & 0.18  & 1.25  & 0.61  & 0.68  & 0.40  \\ \cline{3-8} 
                                                                             &                       & $\mathcal{C_A}$ & 97.77 & 99.45 & 96.8  & 84.29 & 94.03 \\ \cline{3-8} 
                                                                             &                       & $\mathcal{C_B}$ & 83.07 & 98.55 & 96.02 & 76.87 & 90.46 \\ \hline
    \multirow{8}{*}{\begin{tabular}[c]{@{}c@{}}Column\\ Striped\end{tabular}} & \multirow{4}{*}{\texttt{FFFF}} & $\mathcal{M_A}$ & 1.68  & 0     & 1.06  & 1.27  & 0.89  \\ \cline{3-8} 
                                                                             &                       & $\mathcal{M_B}$ & 0.67  & 0    & 0.71  & 0.79  & 0.60  \\ \cline{3-8} 
                                                                             &                       & $\mathcal{C_A}$ & 87.4  & ---   & 93.13 & 90.23 & 89.82 \\ \cline{3-8} 
                                                                             &                       & $\mathcal{C_B}$ & 74.8  & ---   & 92.75 & 89.73 & 88.73 \\ \cline{2-8} 
                                                                             & \multirow{4}{*}{\texttt{0000}} & $\mathcal{M_A}$ & 1.37  & 1.18  & 0.32  & 0.37  & 0.22  \\ \cline{3-8} 
                                                                             &                       & $\mathcal{M_B}$ & 0.54  & 0.94  & 0.19  & 0.20  & 0.12  \\ \cline{3-8} 
                                                                             &                       & $\mathcal{C_A}$ & 46.99 & 90.73 & 51.74 & 32.89 & 34.93 \\ \cline{3-8} 
                                                                             &                       & $\mathcal{C_B}$ & 47.51 & 93.23 & 51.25 & 29.09 & 35.42 \\ \hline
    \multirow{8}{*}{\begin{tabular}[c]{@{}c@{}}Checker-\\ board\end{tabular}} &
      \multirow{4}{*}{\texttt{FFFF}} &
      $\mathcal{M_A}$ &
      1.59 &
      3.05 &
      0.71 &
      0.84 &
      0.61 \\ \cline{3-8} 
                                                                             &                       & $\mathcal{M_B}$ & 0.60  & 1.77  & 0.51  & 0.53  & 0.41  \\ \cline{3-8} 
                                                                             &                       & $\mathcal{C_A}$ & 76.3  & 96.02 & 83.44 & 74.97 & 79.47 \\ \cline{3-8} 
                                                                             &                       & $\mathcal{C_B}$ & 68.07 & 93.81 & 83.96 & 76.62 & 80.64 \\ \cline{2-8} 
                                                                             & \multirow{4}{*}{\texttt{0000}} & $\mathcal{M_A}$ & 1.25  & 2.96  & 0.64  & 0.75  & 0.49  \\ \cline{3-8} 
                                                                             &                       & $\mathcal{M_B}$ & 0.59  & 1.55  & 0.40  & 0.42  & 0.29  \\ \cline{3-8} 
                                                                             &                       & $\mathcal{C_A}$ & 62.61 & 96.02 & 85.71 & 83.54 & 81.54 \\ \cline{3-8} 
                                                                             &                       & $\mathcal{C_B}$ & 60.43 & 93.62 & 83.97 & 82.49 & 79.89 \\ \hline
    \multirow{4}{*}{\begin{tabular}[c]{@{}c@{}}Any\\ Pattern\footnotemark\end{tabular}} &
      \multirow{4}{*}{\begin{tabular}[c]{@{}c@{}}\texttt{5555} /\\ \texttt{AAAA}\end{tabular}} &
      $\mathcal{M_A}$ &
      0 &
      0 &
      0 &
      0 &
      0 \\ \cline{3-8} 
                                                                             &                       & $\mathcal{M_B}$ & 0     & 0     & 0     & 0     & 0     \\ \cline{3-8} 
                                                                             &                       & $\mathcal{C_A}$ & ----  & ----  & ----  & ---   & ---   \\ \cline{3-8} 
                                                                             &                       & $\mathcal{C_B}$ & ----  & ----  & ----  & ---   & ---   \\ \hline
    \multicolumn{2}{|c|}{\multirow{4}{*}{Random}}                                                    & $\mathcal{\mathcal{M_A}}$ & 0.27  & 0.28  & 0.22  & 0.19  & 0.16  \\ \cline{3-8} 
    \multicolumn{2}{|c|}{}                                                                           & $\mathcal{M_B}$ & 0.13  & 0.13  & 0.11  & 0.42  & 0.08  \\ \cline{3-8} 
    \multicolumn{2}{|c|}{}                                                                           & $\mathcal{C_A}$ & 34.83 & 68.94 & 13.97 & 10.18 & 10.12 \\ \cline{3-8} 
    \multicolumn{2}{|c|}{}                                                                           & $\mathcal{C_B}$ & 20.99 & 62.47 & 11.5  & 7.55  & 8.02  \\ \hline
    \multicolumn{8}{l}{$^{\mathrm{*}}$NB. —- not performed as no error occurred at reduced $t_W$ for corr-} \\
    \multicolumn{8}{l}{ esponding data patterns.}
    \end{tabular}
}
\label{Tab:Char1}
\vspace{-0.6cm}
\end{table}
\footnotetext[1]{C1: MR0A16AYS35, C2: MR1A16AYS35, C3\&C4: MR2A16ACYS35, C5: MR2A16AYS35. C1,C2\&C5: Commercial Grade (Temp. range [$0^{\circ}C~70^{\circ}C$]); C3\&C4: Industrial Grade (Temp. range [$-40^{\circ}C~85^{\circ}C$]).}
\footnotetext[2]{Solid/ Striped/ Checkerboard.}

Note that taking the union of the errors resulting from each data pattern (\texttt{0xFFFF, 0xAAAA,0x5555, 0x0000}) can further improve the fault coverage for the worst case. However, only a marginal improvement is observed for the chips used for experiments using different data patterns over the \textit{solid} \texttt{0x0000} data pattern. Besides, Table \ref{Tab:Char1} shows that a vast portion of the erroneous addresses ($\mathcal{M_A}$) generated from different data patterns belongs to those erroneous addresses ($\mathcal{E_A}$) obtained through the address selection strategy from error characterization using \textit{solid} \texttt{0x0000} data pattern (described in strategy 1 of Algo. \ref{alg:mramChar}). Furthermore, it is essential to note that the acquired erroneous addresses heavily depend on the training data and frequency characteristics. Hence, it becomes imperative to select the training data and frequency judiciously.

%------------------------------------------------------------------------------------------%
\vspace{-0.3cm}
\subsection{Evaluation}\label{subsec:res_C}
\begin{figure}[ht!]
\vspace{-0.4cm}
    \centering
    \captionsetup{justification=centering, margin= 0cm}
    \includegraphics[trim=0cm 12.65cm 1.5cm 0cm, clip, width = 0.49\textwidth]{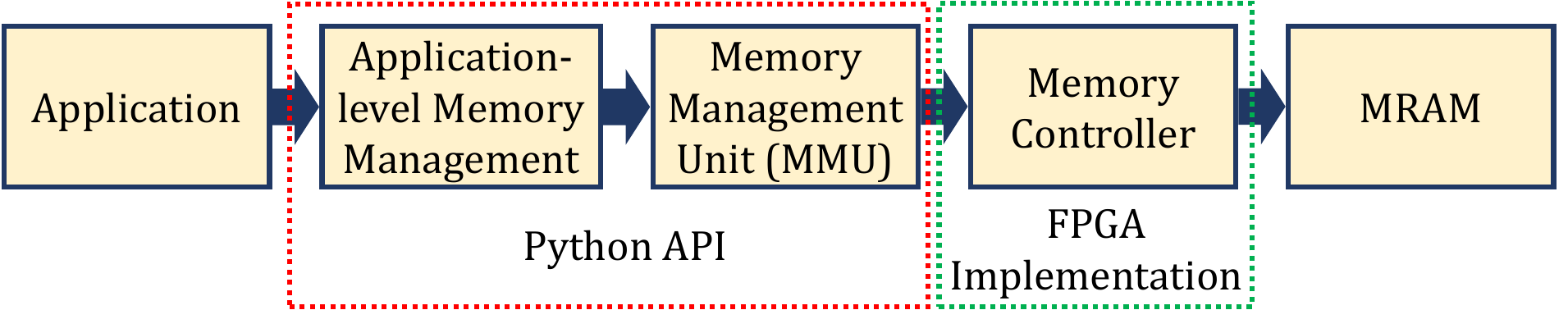}
    \vspace{-0.3cm}
    \caption{Evaluation scheme for proposed AC.}
    \label{fig:Framework}
\vspace{-0.3cm}
\end{figure}

%------------------------------------------------------------------------------------------%
We evaluate our proposed AC framework with two different end-user applications- (i) JPEG image compression and (ii) optical character recognition (OCR) using the K-nearest neighbors (KNN) algorithm \cite{AC_STT_Roy}. Fig. \ref{fig:Framework} presents the evaluation scheme for our proposed AC framework. We use this scheme to emulate MRAM as the main memory for both applications. Conventionally, memory allocation/deallocation (resource management) is accomplished at the application level with the assistance of the operating system (application-level memory management) \cite{pai2012fast}. On the other hand, a memory management unit (MMU) is a separate entity responsible for translating virtual addresses into physical addresses. Although traditionally, MMU is a hardware unit, software implementation of memory management is also feasible without requiring virtual addresses\cite{zagieboylo2020cost}.
%As the traditional hardware-based MMU is not compatible with current toggle MRAM technology \footnote{Currently, commercially available toggle MRAM chips do not support many standard operations of traditional DRAM-based main memory, such as burst mode read/write operation, page mode read, etc.}
To simplify our experimental setup, we fuse the resource manager and the MMU in a single python API (similar to \cite{zagieboylo2020cost}). This python API also interfaces our custom memory controller, built on an FPGA board.

For JPEG application, we use the most common and popular metrics, signal-to-noise ratio (SNR), mean squared error (MSE), and peak signal-to-noise ratio (PSNR), between the approximated image and the unapproximated image used in the JPEG compression to quantify the acceptability of the application output quality. The SNR, MSE, and PSNR are calculated using the following equations (Eqs. \ref{eqn:snr} to \ref{eqn:psnr}). %\st{based on the original image, $f(i,j)$, and the approximated image, $\hat{f}(i,j)$, obtained after performing the proposed approximate framework.}
Here, $\hat{f}(i,j)$ and $f(i,j)$ are the compressed JPEG image data with and without applying our AC framework, respectively. Note that the size of the images is $M \times N$, where $M = N = 90$. Besides, $MAX_I$ is the maximum possible pixel value of the image. Here, the pixels are represented using 8 bits per sample; hence, $MAX_I = 255$.   
%
%\vspace{-0.3cm}
\begin{equation}\label{eqn:snr}
    \begin{split}
    SNR = \frac{\sum_{i = 1}^{M}\sum_{j = 1}^{N}\hat{f}(i,j)^{2}} 
    {\sum_{i = 1}^{M}\sum_{j = 1}^{N}[\hat{f}(i,j) - f(i,j)]^2}
    \end{split}
\end{equation}
\begin{equation}\label{eqn:mse}
    MSE = \frac{1}{MN}\sum_{i = 1}^{M}\sum_{j = 1}^{N}[\hat{f}(i,j) - f(i,j)]^2
\end{equation}
\begin{equation}\label{eqn:psnr}
    PSNR = 10.\log_{10}(\frac{MAX_{I}^2}{MSE})
\end{equation}

\begin{figure*}[ht!]
    \centering
    \captionsetup{justification=centering, margin= 0cm}
    \begin{subfigure}[t]{0.5\textwidth}
        \centering
        \includegraphics[trim=0cm 5cm 10cm 0cm, clip, width=0.99\textwidth]{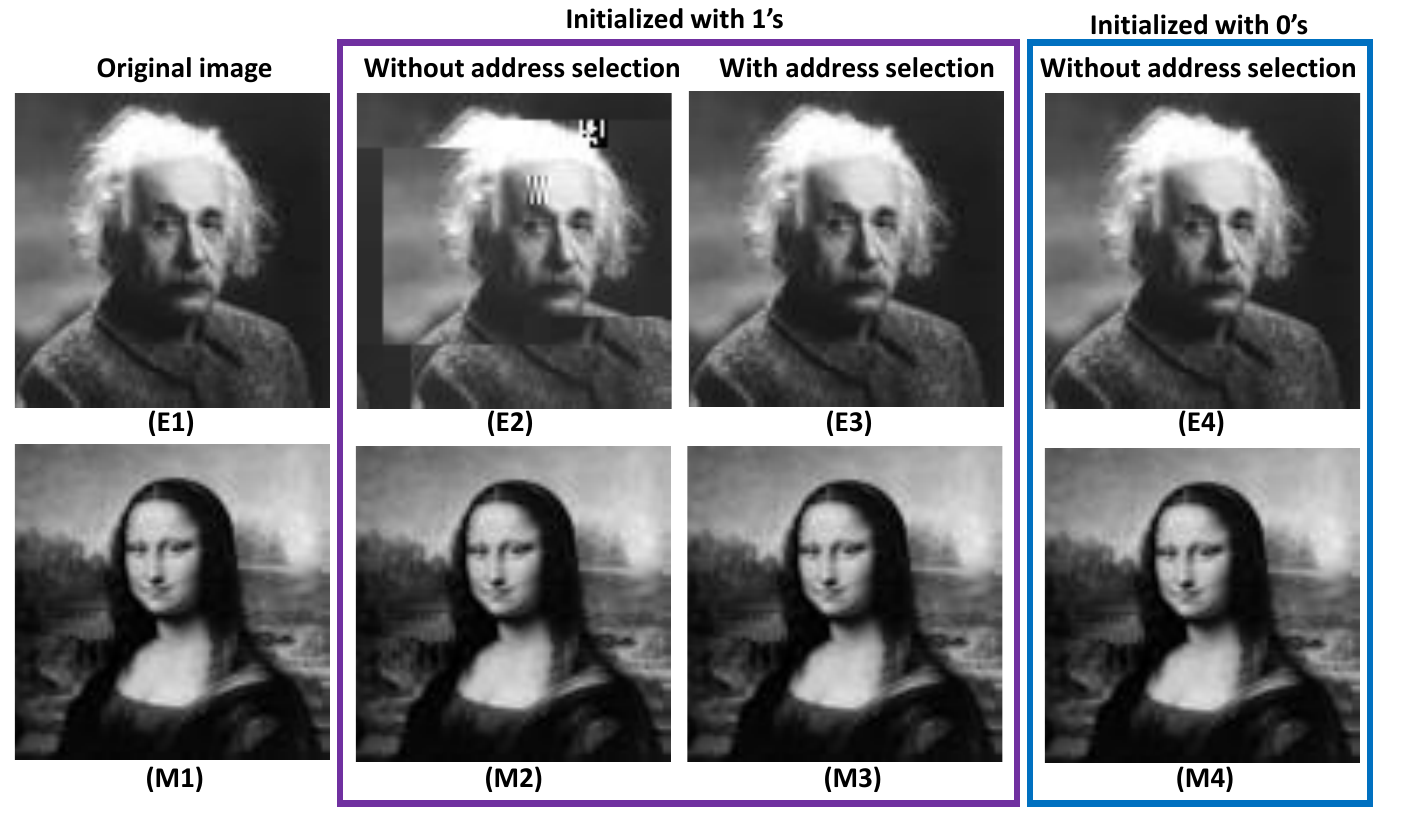}
        \vspace{-0.3cm}
        \caption{}
        \label{fig:mr3ch1}
    \end{subfigure}%
    \begin{subfigure}[t]{0.5\textwidth}
        \centering
        \includegraphics[trim=0cm 5cm 10cm 0cm, clip, width = 0.99\textwidth]{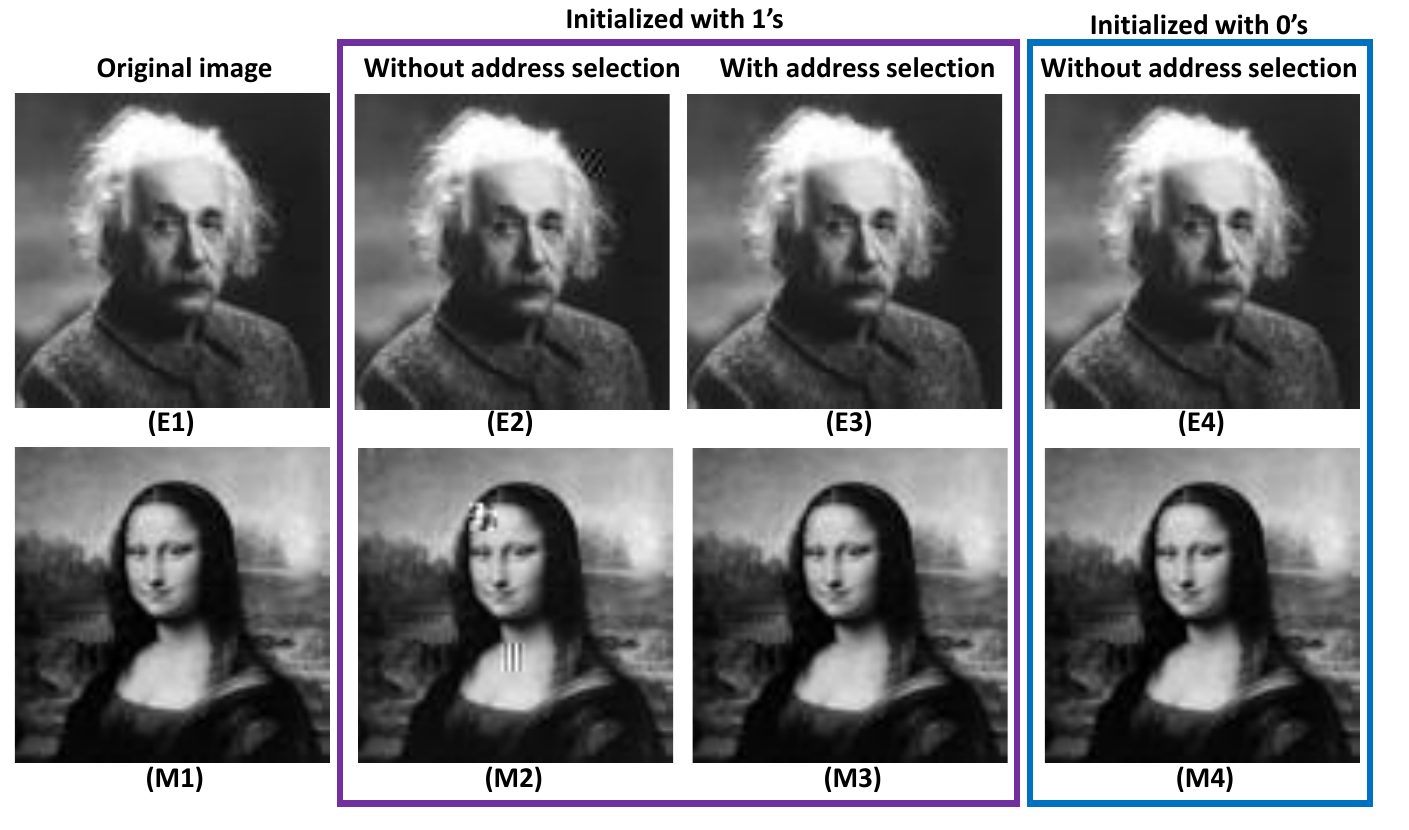}
        \vspace{-0.3cm}
        \caption{}
        \label{fig:mr4ch2}
    \end{subfigure}
    \vspace{-0.4cm}
    \caption{(a) Commercial (b) industrial-grade MRAM results. \nth{1} column of (a) and (b) represents: unapproximated image (90 × 90); \nth{2} \& \nth{3} columns represent: the entire memory is initialized with all 1's before writing the compressed image at reduced $t_W$ into the memory then reading back the memory contents without and with address selection respectively; \nth{4} column represents: the entire memory is initialized with all 0's before writing the compressed image at reduced $t_W$ then reading back the memory contents without address selection.}
    \label{fig:img_proc}
    \vspace{-0.6cm}
\end{figure*}

The value of SNR, MSE, and PSNR measures the image's quality of JPEG compression output, evaluating the proposed scheme's effectiveness. A minimum SNR value is required for such applications for the approximated image where the human brain and eyes can differentiate between the approximated and unapproximated image. A lower SNR 
% \textcolor{green}{(why MSE is inside the bracket? Please make a statement on how SNR and MSE are related.)}
value is considered as unacceptable output quality and vice versa. Similarly, a higher MSE value is considered as unacceptable output quality. The SNR, MSE, and PSNR measurements reveal the degree to which an application could produce satisfactory and reasonable output at reduced $t_W$ for approximate applications and determine the achievable limits of the performance and power improvements for MRAM technology. Fig. \ref{fig:img_proc} shows the unapproximate and approximate output image quality
% (represented quantitatively by SNR and MSE) 
with and without applying the proposed addresses selection technique (strategy 1 of Algo. \ref{alg:mramChar}), considering two different initialization states. The SNR of the image is infinite (MSE of the image is zero) means it does not have any errors relative to the unapproximated image output (Table \ref{Tab:app}). Observing the images and the related SNR and MSE reveals the appropriateness of the proposed technique through application quality. %\st{Note that the compressed test image data are allocated in entirety to each of the required memory addresses.}

%

%Moreover, we evaluate the effectiveness of applying the address selection technique that was discussed in Sec \textcolor{green}{(please refer the section where you discussed the address selection technique)}. 

We have presented our results with and without the address selection technique in this work. In the ``without address selection" technique, our proposed framework produces the worst possible result, where we did not consider erroneous and error-free addresses differently, i.e., memory is allocated conventionally. On the other hand, in the ``with address selection" technique, we excluded error-prone addresses during memory allocation (strategy 1 of Algo. \ref{alg:mramChar}). Note that we did not perform a detailed analysis using sorted addresses (obtained using strategy 2 of Algo. \ref{alg:mramChar}) as reducing $t_{W}$ causes at most ${\sim}6\%$ error in a single measurement. This strategy would only be required when the application needs the entire memory address. Therefore, we did not allocate data based on (lower or higher) error locations. Additionally, this strategy will also include a higher system overhead. 

The result shows the influence of address selection on application quality, which is necessary to understand the impact of approximate data placement in memory addresses. This, in turn, is useful to comprehend how application quality changes with initialized values. In Fig. \ref{fig:img_proc}, the \nth{2} \& \nth{3} columns show the JPEG image without and with applying the address selection technique, respectively. It is evident that the image quality improvement obtained using the proposed address selection technique is substantially better than without applying the technique, underscoring the need for the proposed address selection. However, as switching from ``1" $\rightarrow$ ``0" is more vulnerable at reduced $t_W$, requiring more write current for the prolonged duration; hence, initializing the entire memory with all 1's requires address selection technique to attain very small or no loss in the output quality. Contrarily, initializing the entire memory with all 0's does not require an address selection technique to obtain the same level of output quality. Furthermore, we also observe that commercial-grade MRAM is more vulnerable to reduced $t_W$ than industrial-grade MRAM as commercial-grade MRAM is more susceptible to thermal noise (see Sect. \ref{subsec:res_D}). Therefore, without applying the address selection technique in industrial-grade MRAMs, we notice a negligible loss in the output quality. It is worth mentioning that JPEG compression is a critical application in approximate computing. For example, a slight change in the RUNLENGTH (number of consecutive zero-valued discrete cosine transform (DCT) coefficient of the compressed JPEG data \cite{wallace1992jpeg}) can impact the output quality severely.

In addition to JPEG image compression, we also demonstrate our AC framework for 
% \textcolor{green}{(using or for? In my understanding, it's for.)} 
the KNN algorithm. KNN algorithm is a widely used supervised machine learning technique for image classification and object recognition (detail construction of the algorithm is out of this paper's scope). In this work, we use the UCI optical character recognition (OCR) dataset \cite{OCR} and the KNN algorithm (vanilla). To evaluate the performance of our proposed framework, we compare KNN accuracy with and without using our AC framework. Accuracy quantifies the fraction of correct prediction out of all test samples. The accuracy of the KNN algorithm without using our AC framework is $95.5\%$. If the MRAM is initialized with logic ``0" bits, our AC framework does not impact the KNN accuracy. On the other hand, with the logic ``1" initialization and without address selection technique, the accuracy of the KNN algorithm using our AC framework ranges between $88.25\%$ to $92.25\%$ (i.e., ${<}7.6\%$ performance degradation). With the logic ``1" initialization and address selection technique, however, the accuracy of the KNN algorithm with our frameworks improves to at least $92.5\%$ (i.e., ${<}3.2\%$ performance degradation).

\vspace{-0.3cm}
%------------------------------------------------------------------------------------------%
\subsection{Effect of Operating Condition and Chip Variations}\label{subsec:res_D}
%------------------------------------------------------------------------------------------%

Random thermal fluctuations and process variation exacerbate the intrinsic stochasticity of the magnetic orientation switching of MRAM cells; hence, significantly varies the switching time of the cell over the entire memory array. Therefore, the MRAM write failure rate increases significantly as temperature increases \cite{Ferdaus2021}. We can characterize these failures at the highest possible operating temperature to incorporate the temperature effect to cover maximum erroneous addresses. 
%\st{compared with the maximum operating temperature used during normal execution}. 
Our observation is that errors occurring at a lower temperature are a subset of those occurring at high temperatures; therefore, guaranteeing no additional bit errors in the lower temperature range. The highest temperature used for characterization can be determined within the manufacturer's recommended value covering all MRAM models (in our case, $65^{\circ}C$), suitable for commercial applications. Our experiments' characterization and evaluation are performed at room temperature ($26^{\circ}C$). To demonstrate the temperature effect on MRAM errors, we evaluate results for both high ($65^{\circ}C$) and low ($20^{\circ}C$) temperatures and compare them with room temperature ($26^{\circ}C$).%\st{ The temperature range is chosen within the manufacturer's recommended value, which is $[0^{\circ}C~70^{\circ}C]$ suitable for all memory models used in the experiment. }

We observe that the total number of errors and erroneous addresses increase with higher temperatures. However, no significant change is observed in the total number of erroneous addresses at low temperature ($20^{\circ}C$) and temperature range up to $45^{\circ}C$. As the thermal noise becomes smaller at low temperature, the low temperature has less impact at reduced $t_{W}$. Therefore, only results at $65^{\circ}C$ temperature (worst case) are shown in Table \ref{Tab:Char_ht} as the obtained errors are insignificant at low temperatures. Moreover, the maximum erroneous address ($\mathcal{M_A}$) at $65^{\circ}C$ in reduced $t_W$ ($5ns$) is $16.00\%$ for any given pattern (except \texttt{0xAAAA} or \texttt{0x5555}), considering a single test measurement for all memory chips. In contrast, the minimum is only $0.002\%$. Similarly, the maximum erroneous bit ($\mathcal{M_B}$) is $8.98\%$, whereas the minimum is only $0.002\%$.
% The obtained erroneous addresses ($\mathcal{M_A}$) (and bits ($\mathcal{M_B}$)) at $65^{\circ}C$ in reduced $t_W$ 
% \textcolor{green}{(again for all reduced tws?)}
% (= $5ns$), are in the range $[0.002\%~16.00\%]~([0.002\%~8.98\%])$ for any data patterns (except \texttt{0xAAAA} or \texttt{0x5555}) considering all memory chips. 
Comparing with the nominal operating condition (Table \ref{Tab:Char1}), we observe a maximum of $\sim$$1.65\times$ increase in the erroneous addresses and $\sim$$1.72\times$ increase in the erroneous bits at $65^{\circ}C$ temperature. Contrarily, the maximum erroneous address ($\mathcal{M_A}$) at $20^{\circ}C$ in reduced $t_W$ ($5ns$) is $3.89\%$. In contrast, the minimum is only $0.0004\%$. Similarly, the maximum erroneous bit ($\mathcal{M_B}$) is $2.21\%$, whereas the minimum is only $0.0001\%$.
% Contrarily, the obtained erroneous addresses ($\mathcal{M_A}$) (and bits ($\mathcal{M_B}$)) at $20^{\circ}C$ in reduced $t_W$
% \textcolor{green}{(again for all reduced tws?)} 
% (= $5ns$) are in the range $[0.0004\%~3.89\%]~([0.0001\%~2.21\%])$ for any data patterns considering all memory chips. 
Comparing with the nominal operating condition (Table \ref{Tab:Char1}), we observe a maximum of $\sim$$0.36\times$ decrease in the erroneous addresses and $\sim$$0.33\times$ decrease in the erroneous bits at $20^{\circ}C$ temperature. Although a detailed assessment is not performed, the obtained results reveal that errors at lower temperatures are a subset of the errors at high temperatures.  

Furthermore, to observe the impact of external magnetic field (M-Field), we collected a set of test data sequences at room temperature ($26^{\circ}C$) with an ${\sim}8mT$ external M-Field by %\st{applying a constant rare earth magnetic source generated from} 
using a set of permanent magnets in six different orientations of 3D coordinates\footnote{The maximum recommended magnetic field is ${\sim}10mT$ for commercial-grade and ${\sim}12.5mT$ for industrial-grade MRAM.}. However, no significant change is observed in the number of erroneous cells with ($\sim 8mT$) external M-Field.
The maximum erroneous address ($\mathcal{M_A}$) in reduced $t_W$ ($5ns$) is $7.60\%$. In contrast, the minimum is only $0.005\%$. Similarly, the maximum erroneous bit ($\mathcal{M_B}$) is $4.52\%$, whereas the minimum is only $0.0002\%$.
% The obtained erroneous addresses ($\mathcal{M_A}$) ( and bits ($\mathcal{M_B}$)) at the reduced $t_W$ (= $5ns$)
% \textcolor{green}{(for any specific value of reduced tw or at any reduced tw?)}
% are in the range $[0.005\%~7.60\%]~([0.0002\%~4.52\%])$ for any data patterns considering all memory chips. 
Comparing with the nominal operating condition (Table \ref{Tab:Char1}), we observe a maximum of $\sim$$0.26\times$ increase in the erroneous addresses and $\sim$$0.37\times$ increase in the erroneous bits. Hence, we can conclude that the external M-Field does not significantly impact the number of error occurrences.  

The silicon results from five different memory models of three different sizes show that the statistics of erroneous cells are different for different memory models, shown in Tables \ref{Tab:Char1} and \ref{Tab:Char_ht}. These sources of variations come from architectural as well as both inter- and intra-chip dissimilarities due to the random process variation, which is the key source of any memory chips' randomness. Data collected from different FPGAs also verify that the memory controllers do not influence the proposed framework.

\begin{table}[ht!]
\caption{Error Statistics at High Temperature.}
\setcellgapes{1pt}%parameter for the spacing
\captionsetup{justification=centering, margin= 0cm}
%\vspace{-0.3cm}
\makegapedcells
\centering
\setlength\tabcolsep{5pt} 
\resizebox{0.48\textwidth}{!}
{
   \begin{tabular}{|c|c|c|c|c|c|c|c|}
    \hline
    \multicolumn{2}{|c|}{Sample Chip} & (\%) & C1    & C2    & C3    & C4    & C5    \\ \hline
    \multirow{8}{*}{Solid} &
      \multirow{4}{*}{\texttt{FFFF}} &
      $\mathcal{M_A}$ &
      0.002 &
      0.09 &
      0 &
      0 &
      0 \\ \cline{3-8} 
          &                           & $\mathcal{M_B}$ & 0.002 & 0.02  & 0     & 0     & 0     \\ \cline{3-8} 
          &                           & $\mathcal{C_A}$ & 100   & 5.79  & ----  & ----  & ----  \\ \cline{3-8} 
          &                           & $\mathcal{C_B}$ & 100   & 16.77 & ----  & ----  & ----  \\ \cline{2-8} 
          & \multirow{4}{*}{\texttt{0000}}     & $\mathcal{M_A}$ & 16.00 & 13.53 & 7.55  & 2.73  & 0.94  \\ \cline{3-8} 
          &                           & $\mathcal{M_B}$ & 6.31  & 7.54  & 4.30  & 0.56  & 0.22  \\ \cline{3-8} 
          &                           & $\mathcal{C_A}$ & 39.70 & 77.63 & 10.72 & 11.77 & 21.8  \\ \cline{3-8} 
          &                           & $\mathcal{C_B}$ & 26.59 & 74.68 & 7.93  & 17.07 & 32.54 \\ \hline
    \multirow{8}{*}{\begin{tabular}[c]{@{}c@{}}Row\\ Striped\end{tabular}} &
      \multirow{4}{*}{\texttt{FFFF}} &
      $\mathcal{M_A}$ &
      12.83 &
      6.92 &
      4.63 &
      1.61 &
      0.33 \\ \cline{3-8} 
          &                           & $\mathcal{M_B}$ & 6.05  & 3.87  & 1.85  & 0.35  & 0.12  \\ \cline{3-8} 
          &                           & $\mathcal{C_A}$ & 47.39 & 81.38 & 13.84 & 15.64 & 38.17 \\ \cline{3-8} 
          &                           & $\mathcal{C_B}$ & 32.07 & 79.25 & 13.31 & 21.56 & 42.08 \\ \cline{2-8} 
          & \multirow{4}{*}{\texttt{0000}}     & $\mathcal{M_A}$ & 12.79 & 5.74  & 2.87  & 1.01  & 0.167 \\ \cline{3-8} 
          &                           & $\mathcal{M_B}$ & 5.61  & 3.07  & 1.33  & 0.21  & 0.07  \\ \cline{3-8} 
          &                           & $\mathcal{C_A}$ & 30.32 & 78.44 & 6.74  & 5.91  & 14.65 \\ \cline{3-8} 
          &                           & $\mathcal{C_B}$ & 18.65 & 73.35 & 3.11  & 9.61  & 22.77 \\ \hline
    \multirow{8}{*}{\begin{tabular}[c]{@{}c@{}}Column\\ Striped\end{tabular}} &
      \multirow{4}{*}{\texttt{FFFF}} &
      $\mathcal{M_A}$ &
      15.56 &
      9.93 &
      3.55 &
      1.02 &
      0.273 \\ \cline{3-8} 
          &                           & $\mathcal{M_B}$ & 8.98  & 5.41  & 1.04  & 0.24  & 0.12  \\ \cline{3-8} 
          &                           & $\mathcal{C_A}$ & 47.66 & 74.21 & 12.32 & 16.07 & 48.42 \\ \cline{3-8} 
          &                           & $\mathcal{C_B}$ & 27.41 & 68.75 & 15.01 & 23.16 & 53.61 \\ \cline{2-8} 
          & \multirow{4}{*}{\texttt{0000}}     & $\mathcal{M_A}$ & 3.15  & 2.13  & 2.55  & 0.83  & 0.23  \\ \cline{3-8} 
          &                           & $\mathcal{M_B}$ & 0.76  & 0.93  & 0.68  & 0.15  & 0.09  \\ \cline{3-8} 
          &                           & $\mathcal{C_A}$ & 28.25 & 95.81 & 9.32  & 14.93 & 36.38 \\ \cline{3-8} 
          &                           & $\mathcal{C_B}$ & 41.57 & 94.11 & 14.83 & 23.5  & 37.36 \\ \hline
    \multirow{8}{*}{\begin{tabular}[c]{@{}c@{}}Checker-\\ board\end{tabular}} &
      \multirow{4}{*}{\texttt{FFFF}} &
      $\mathcal{M_A}$ &
      9.40 &
      5.93 &
      3.34 &
      0.84 &
      0.23 \\ \cline{3-8} 
          &                           & $\mathcal{M_B}$ & 4.75  & 3.21  & 0.86  & 0.20  & 0.11  \\ \cline{3-8} 
          &                           & $\mathcal{C_A}$ & 50.53 & 80.52 & 12.11 & 17.01 & 48.17 \\ \cline{3-8} 
          &                           & $\mathcal{C_B}$ & 33.8  & 75.31 & 16.94 & 26.46 & 53.5  \\ \cline{2-8} 
          & \multirow{4}{*}{\texttt{0000}}     & $\mathcal{M_A}$ & 9.10  & 5.96  & 3.37  & 0.89  & 0.25  \\ \cline{3-8} 
          &                           & $\mathcal{M_B}$ & 4.37  & 3.06  & 0.93  & 0.19  & 0.11  \\ \cline{3-8} 
          &                           & $\mathcal{C_A}$ & 40.51 & 78.21 & 8.82  & 14.67 & 41.71 \\ \cline{3-8} 
          &                           & $\mathcal{C_B}$ & 24.48 & 71.84 & 12.77 & 22.76 & 41.62 \\ \hline
    \multirow{4}{*}{\begin{tabular}[c]{@{}c@{}}Any\\ Pattern\end{tabular}} &
      \multirow{4}{*}{\begin{tabular}[c]{@{}c@{}}\texttt{5555}/\\ \texttt{AAAA}\end{tabular}} &
      $\mathcal{M_A}$ &
      0 &
      0 &
      0 &
      0 &
      0 \\ \cline{3-8} 
          &                           & $\mathcal{M_B}$ & 0     & 0     & 0     & 0     & 0     \\ \cline{3-8} 
          &                           & $\mathcal{C_A}$ & ----  & ----  & ----  & ----  & ----  \\ \cline{3-8} 
          &                           & $\mathcal{C_B}$ & ----  & ----  & ----  & ----  & ----  \\ \hline
    \multicolumn{2}{|c|}{\multirow{4}{*}{Random}} &
      $\mathcal{M_A}$ &
      0.13 &
      0.23 &
      0.01 &
      0.005 &
      0.008 \\ \cline{3-8} 
    \multicolumn{2}{|c|}{}            & $\mathcal{M_B}$ & 0.04  & 0.09  & 0.01  & 0.002 & 0.002 \\ \cline{3-8} 
    \multicolumn{2}{|c|}{}            & $\mathcal{C_A}$ & 59.09 & 72.97 & 31.43 & 38.46 & 38.1  \\ \cline{3-8} 
    \multicolumn{2}{|c|}{}            & $\mathcal{C_B}$ & 33.16 & 63.81 & 36.3  & 40.48 & 41.58 \\ \hline
    \multicolumn{8}{l}{$^{\mathrm{*}}$NB. —- not performed as no error occurred at reduced $t_W$ for corr-} \\
    \multicolumn{8}{l}{ esponding data patterns.} \\
    \end{tabular}
}
\label{Tab:Char_ht}
\vspace{-0.5cm}
\end{table}

\begin{figure*}[ht!]
    \centering
    \captionsetup{justification=centering, margin= 0.5cm}
    \begin{subfigure}[t]{0.4\textwidth}
        \centering
        \includegraphics[trim=0cm 5cm 16cm 0cm, clip, width=0.99\textwidth]{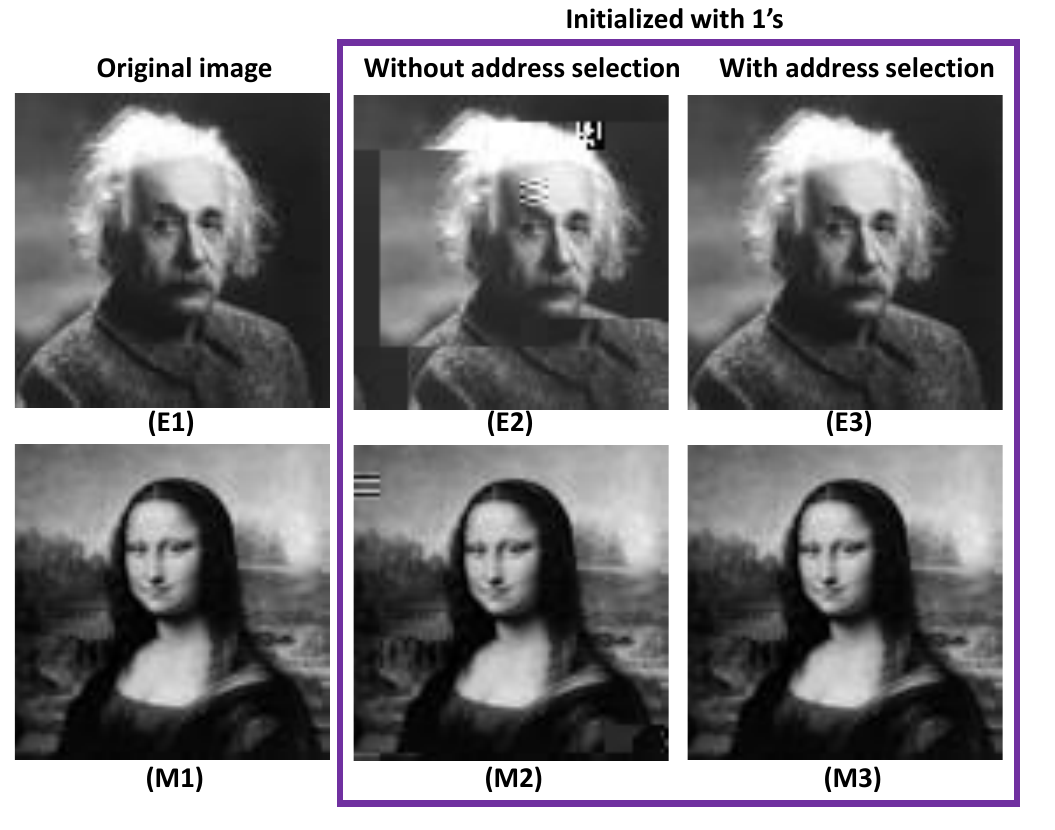}
        \vspace{-0.5cm}
        \caption{}
        \label{fig:mr3ch1_ht}
    \end{subfigure}%
    \begin{subfigure}[t]{0.4\textwidth}
        \centering
        \includegraphics[trim=0cm 5cm 16cm 0cm, clip, width = 0.99\textwidth]{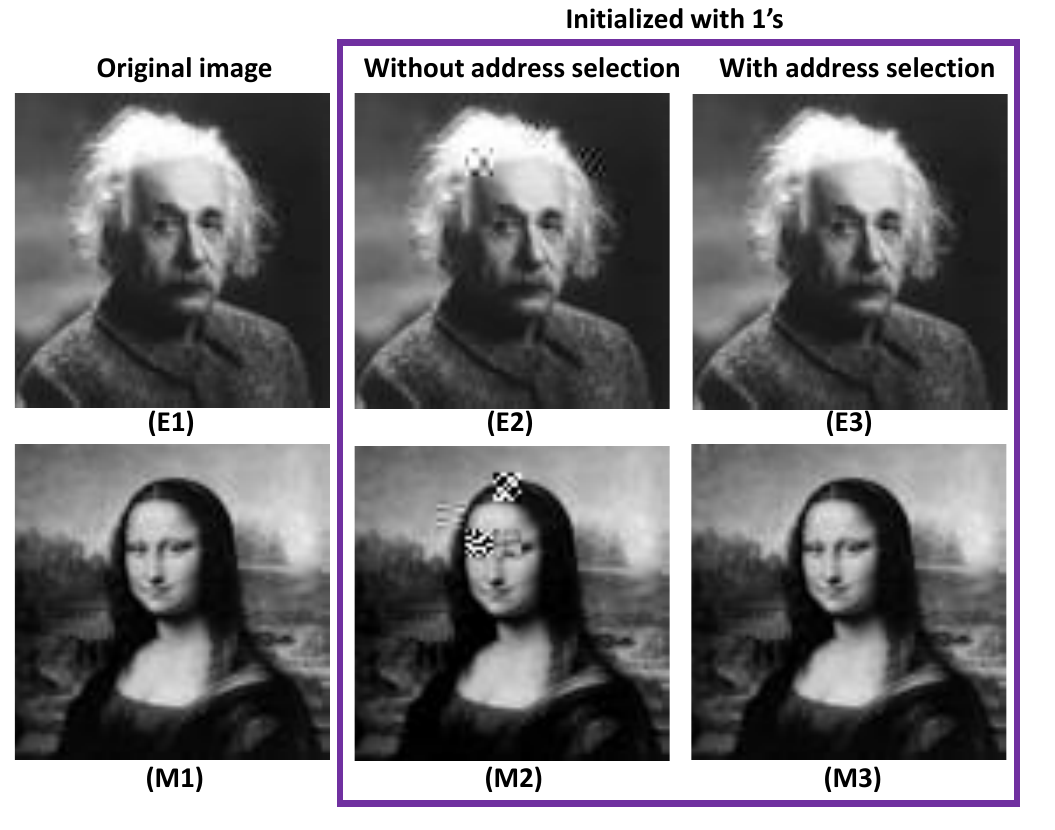}
        \vspace{-0.5cm}
        \caption{}
        \label{fig:mr4ch2_ht}
    \end{subfigure}
    \vspace{-0.3cm}
    \caption{(a) Commercial (b) industrial-grade MRAM results. \nth{1} column of (a) and (b) represents the unapproximated image (90 × 90); \nth{2} \& \nth{3} columns represent: the entire memory is initialized with all 1's before writing the compressed image at reduced $t_W$ into the memory then reading back the memory contents without and with address selection respectively at $65^{\circ}C$.}
    % \footnotemark}
    \label{fig:img_proc_ht}
\vspace{-0.2cm}
\end{figure*}

Furthermore, we have also evaluated our result for both JPEG compression and the KNN algorithm at high temperature ($65^{\circ}C$). Fig. \ref{fig:img_proc_ht} presents the JPEG images at $65^{\circ}C$ with ``1" initialization (initialization ``0" has no impact). Compared with Fig. \ref{fig:img_proc}, the image quality is slightly worse (noise magnitude increased by $2{\times}$).
% \textcolor{green}{(who are you comparing with?)}. 
On the other hand, the KNN accuracy further decreased by $3.2\%$ (at most). Table \ref{Tab:app} summarizes the list of applications, datasets, and results at both room temperature ($26^{\circ}C$) and high temperature ($65^{\circ}C$).

% \textcolor{blue}{Furthermore, Table \ref{Tab:app} lists the applications, datasets, and results used to evaluate the proposed framework (strategy 1 of Algorithm 1). We evaluate our proposed framework using two popular recognition and classification applications, listed in Table \ref{Tab:app}. For each application, the table also lists the underlying algorithm, the data sets, quality metrics used to evaluate the output quality, and the worst-case (i.e., from multiple similar test categories, the worst one is exhibited) results considering different memory grades and extreme operating conditions. We utilized classification accuracy, i.e., the fraction of input samples (feature vectors) correctly recognized or classified, as an estimate for output quality of the optical character recognition (OCR) of handwritten digits data set (from a total of 43 people, 30 contributed to the training set and different 13 to the test set) using K-nearest neighbors (KNN) algorithm\footnote{If reviewers ask we can also present the confusion matrix of the classifier in the final version of the paper} \cite{OCR}. Note that each element of the $8 \times 8$ input matrix (64 attributes) of the OCR data set is an integer of range $[0~16]$. Hence during the data pre-processing phase, we transform the integers of the input matrix in the range of $[0~15]$ to manipulate as hex digits and store the input matrix in the memory at reduced $t_W$. On the other hand, during JPEG compression, only compressed image data is stored in the memory at reduced $t_W$.}

\begin{table*}[ht!]
\caption{Application benchmarks and the worst-case results.}
\setcellgapes{1pt}%parameter for the spacing
\captionsetup{justification=centering, margin= 0cm}
% \vspace{-0.3cm}
\makegapedcells
\centering
\setlength\tabcolsep{1pt} 
\resizebox{0.99\textwidth}{!}
{
\begin{tabular}{cccccccccccccccccccc}
\hline
\multicolumn{4}{|c|}{Chip grade} & \multicolumn{8}{c|}{Commerical} & \multicolumn{8}{c|}{Industrial} \\ \hline
\multicolumn{4}{|c|}{Operating condition} & \multicolumn{4}{c|}{$T_{Room}$  ($26^{\circ}C$)} & \multicolumn{4}{c|}{$T_{High}$ ($65^{\circ}C$)} & \multicolumn{4}{c|}{$T_{Room}$  ($26^{\circ}C$)} & \multicolumn{4}{c|}{$T_{High}$ ($65^{\circ}C$)} \\ \hline
\multicolumn{4}{|c|}{Initial state} & \multicolumn{2}{c|}{``1"} & \multicolumn{2}{c|}{``0"} & \multicolumn{2}{c|}{``1"} & \multicolumn{2}{c|}{``0"} & \multicolumn{2}{c|}{``1"} & \multicolumn{2}{c|}{``0"} & \multicolumn{2}{c|}{``1"} & \multicolumn{2}{c|}{``0"} \\ \hline
\multicolumn{1}{|c|}{\multirow{2}{*}{App.}} & \multicolumn{1}{c|}{\multirow{2}{*}{Algorithm}} & \multicolumn{1}{c|}{\multirow{2}{*}{\begin{tabular}[c]{@{}c@{}}Approx.\\ data\end{tabular}}} & \multicolumn{1}{c|}{\multirow{2}{*}{\begin{tabular}[c]{@{}c@{}}Quality\\ metric\end{tabular}}} & \multicolumn{2}{c|}{Addr. sel.} & \multicolumn{2}{c|}{Addr. sel.} & \multicolumn{2}{c|}{Addr. sel.} & \multicolumn{2}{c|}{Addr. sel.} & \multicolumn{2}{c|}{Addr. sel.} & \multicolumn{2}{c|}{Addr. sel.} & \multicolumn{2}{c|}{Addr. sel.} & \multicolumn{2}{c|}{Addr. sel.} \\ \cline{5-20} 
\multicolumn{1}{|c|}{} & \multicolumn{1}{c|}{} & \multicolumn{1}{c|}{} & \multicolumn{1}{c|}{} & \multicolumn{1}{c|}{W/O} & \multicolumn{1}{c|}{W} & \multicolumn{1}{c|}{W/O} & \multicolumn{1}{c|}{W} & \multicolumn{1}{c|}{W/O} & \multicolumn{1}{c|}{W} & \multicolumn{1}{c|}{W/O} & \multicolumn{1}{c|}{W} & \multicolumn{1}{c|}{W/O} & \multicolumn{1}{c|}{W} & \multicolumn{1}{c|}{W/O} & \multicolumn{1}{c|}{W} & \multicolumn{1}{c|}{W/O} & \multicolumn{1}{c|}{W} & \multicolumn{1}{c|}{W/O} & \multicolumn{1}{c|}{W} \\ \hline
\multicolumn{1}{|c|}{\multirow{3}{*}{\begin{tabular}[c]{@{}c@{}}JPEG\\ comp.\end{tabular}}} & \multicolumn{1}{c|}{\multirow{3}{*}{\begin{tabular}[c]{@{}c@{}}JPEG\\ enc.\end{tabular}}} & \multicolumn{1}{c|}{\multirow{3}{*}{\begin{tabular}[c]{@{}c@{}}JPEG\\ image\end{tabular}}} & \multicolumn{1}{c|}{SNR} & \multicolumn{1}{c|}{4.34} & \multicolumn{1}{c|}{$\infty$} & \multicolumn{1}{c|}{$\infty$} & \multicolumn{1}{c|}{$\infty$} & \multicolumn{1}{c|}{4.35} & \multicolumn{1}{c|}{${>}10^5$} & \multicolumn{1}{c|}{$\infty$} & \multicolumn{1}{c|}{$\infty$} & \multicolumn{1}{c|}{122.00} & \multicolumn{1}{c|}{${>}10^4$} & \multicolumn{1}{c|}{$\infty$} & \multicolumn{1}{c|}{$\infty$} & \multicolumn{1}{c|}{36.99} & \multicolumn{1}{c|}{${>}10^4$} & \multicolumn{1}{c|}{$\infty$} & \multicolumn{1}{c|}{$\infty$} \\ \cline{4-20} 
\multicolumn{1}{|c|}{} & \multicolumn{1}{c|}{} & \multicolumn{1}{c|}{} & \multicolumn{1}{c|}{MSE} & \multicolumn{1}{c|}{3216.11} & \multicolumn{1}{c|}{0.00} & \multicolumn{1}{c|}{0.00} & \multicolumn{1}{c|}{0.00} & \multicolumn{1}{c|}{3208.83} & \multicolumn{1}{c|}{0.06} & \multicolumn{1}{c|}{0.00} & \multicolumn{1}{c|}{0.00} & \multicolumn{1}{c|}{111.52} & \multicolumn{1}{c|}{0.85} & \multicolumn{1}{c|}{0.00} & \multicolumn{1}{c|}{0.00} & \multicolumn{1}{c|}{339.17} & \multicolumn{1}{c|}{1.01} & \multicolumn{1}{c|}{0.00} & \multicolumn{1}{c|}{0.00} \\ \cline{4-20} 
\multicolumn{1}{|c|}{} & \multicolumn{1}{c|}{} & \multicolumn{1}{c|}{} & \multicolumn{1}{c|}{PSNR} & \multicolumn{1}{c|}{13.06} & \multicolumn{1}{c|}{$\infty$} & \multicolumn{1}{c|}{$\infty$} & \multicolumn{1}{c|}{$\infty$} & \multicolumn{1}{c|}{13.07} & \multicolumn{1}{c|}{60.40} & \multicolumn{1}{c|}{$\infty$} & \multicolumn{1}{c|}{$\infty$} & \multicolumn{1}{c|}{27.66} & \multicolumn{1}{c|}{48.85} & \multicolumn{1}{c|}{$\infty$} & \multicolumn{1}{c|}{$\infty$} & \multicolumn{1}{c|}{22.83} & \multicolumn{1}{c|}{48.09} & \multicolumn{1}{c|}{$\infty$} & \multicolumn{1}{c|}{$\infty$} \\ \hline
\multicolumn{1}{|c|}{OCR} & \multicolumn{1}{c|}{KNN} & \multicolumn{1}{c|}{Feat. vec.} & \multicolumn{1}{c|}{\% Acc.*} & \multicolumn{1}{c|}{88.25} & \multicolumn{1}{c|}{92.75} & \multicolumn{1}{c|}{95.5} & \multicolumn{1}{c|}{95.5} & \multicolumn{1}{c|}{85.25} & \multicolumn{1}{c|}{91.75} & \multicolumn{1}{c|}{95.5} & \multicolumn{1}{c|}{95.5} & \multicolumn{1}{c|}{92.25} & \multicolumn{1}{c|}{93.5} & \multicolumn{1}{c|}{95.5} & \multicolumn{1}{c|}{95.5} & \multicolumn{1}{c|}{92.00} & \multicolumn{1}{c|}{92.75} & \multicolumn{1}{c|}{95.5} & \multicolumn{1}{c|}{95.5} \\ \hline
\multicolumn{20}{l}{*KNN Accuracy (Acc.) without approximation = 95.5\%; W/O: without address selection; W: with address selection}
%\vspace{-0.45cm}
\end{tabular}
}
\label{Tab:app}
\vspace{-0.6cm}
\end{table*}

%\vspace{-0.45cm}

% \footnotetext{While working on the image experiment, we do not have access to the mentioned ThermoStream; hence unable to reach $65^{\circ}C$ (used for characterization). In the revised version, we are planning to show image results at $65^{\circ}C$.}
%
\vspace{-0.3cm}
\subsection{Power Improvement} \label{subsubsec:powImpv}
MRAM write energy ($E_{W}$) can be expressed as by Eq. \ref{eqn:pwr}. Here, $V$ is the write pulse magnitude, $I(t)$ is the write current that depends on instantaneous cell resistance ($R_{Cell}$) during the resistance switching process. A higher value of $t_{W}$ will cost a higher $E_{W}$.
\vspace{-0.2cm}
\begin{equation}\label{eqn:pwr}
    E_{W}= V\int_{0}^{t_{W}}I(t)dt
\end{equation}
\vspace{-0.2cm}

To evaluate the energy efficiency of our proposed framework, we require to measure the write current over $t_{W}$. However, it is difficult to achieve an appropriate experimental setup
% \textcolor{green}{(Achieve what?)}
using COTS chips. For example, all of our test chips have multiple pairs of power-ground pins. Therefore, measuring write current requires probing each power/ground current with sufficiently high-frequency bandwidth instruments while preserving precise timestamps. Additionally, power consumed by memory peripherals (I/Os, sense-amplifier, etc.) is hard to estimate. Therefore, we estimate the power savings of our AC scheme from simulation data. For power analysis, we use the simulation framework for Spin-Transfer Torque MRAM (STT-MRAM), developed by Wang \textit{et al.} \cite{wang2016comparative}. 
It is worth mentioning that STT-MRAM is an improved version of MRAM technology,
% \textcolor{green}{(which MRAM? The first generation?)}, 
which is more scalable, hence achieving higher densities at a lower price, providing a significant reduction in switching energy and consuming low power compared to toggle MRAM devices. In STT-MRAM, the current is injected perpendicularly into the magnetic tunnel junction (MTJ), and the read/write operation is performed through the same path \cite{SOT, khvalkovskiy2013basic}. Our simulation result aligns with the error characteristics of toggle MRAM (see Sect. \ref{subsec:res_B}), confirming our proposed AC framework's applicability in next-generation MRAM technology. Note that all simulation parameters in this work are adopted from \cite{wang2016comparative} and \cite{song2018evaluation}, including physical variation (MTJ tunnel thickness, free layer thickness, cell area, etc.) that might occur from the fabrication process.

\begin{figure}[ht!]
%\vspace{-0.3cm}
\centering
\captionsetup{justification=centering, margin= 0cm}
\begin{minipage} []{.24\textwidth}
    \centering
    \captionsetup{justification=centering, margin= 0cm}
    \begin{subfigure}[t]{1\textwidth}
        \centering
        \includegraphics[trim=0.25cm 0.25cm 0.25cm 0.25cm, clip, width=0.98\textwidth]{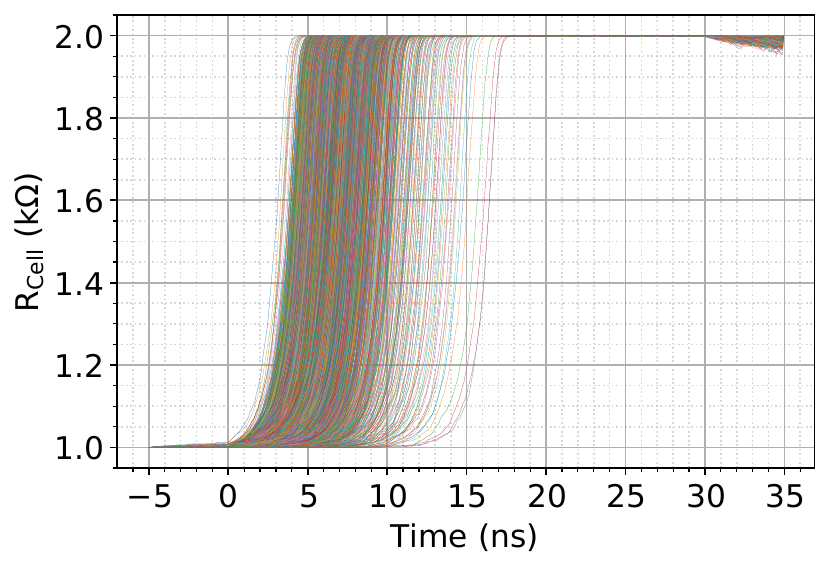}
        \caption{}
        \label{fig:logic0_resistance}
    \end{subfigure}%
    \vspace{\medskipamount}
    \begin{subfigure}[t]{1\textwidth}
        \centering
        \includegraphics[trim=0.25cm 0.25cm 0.25cm 0.25cm, clip, width = 0.98\textwidth]{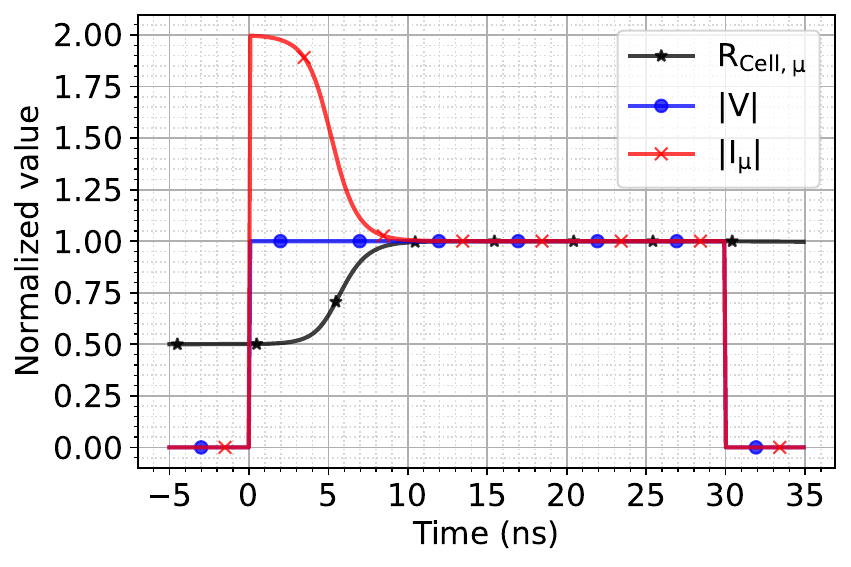}
        \caption{}
        \label{fig:logic0_resistance_voltage_current}
    \end{subfigure}%
    \vspace{\medskipamount}
    \begin{subfigure}[t]{1\textwidth}
        \centering
        \includegraphics[trim=0.25cm 0.25cm 0.25cm 0.25cm, clip, width = 0.98\textwidth]{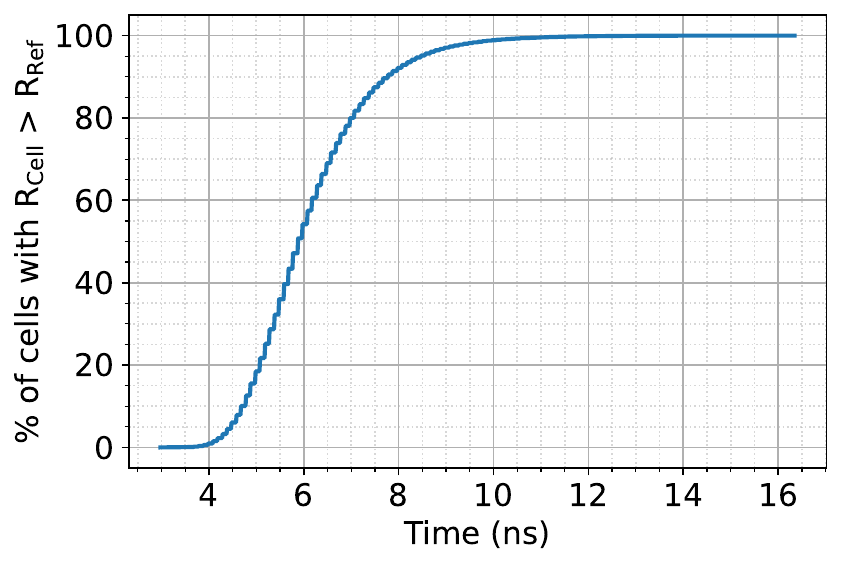}
        \caption{}
        \label{fig:logic0_cdf}
    \end{subfigure}
\end{minipage}
\begin{minipage} []{.24\textwidth}
    \centering
    \captionsetup{justification=centering, margin= 0cm}
    \begin{subfigure}[t]{1\textwidth}
        \centering
        \includegraphics[trim=0.25cm 0.25cm 0.25cm 0.25cm, clip, width=0.98\textwidth]{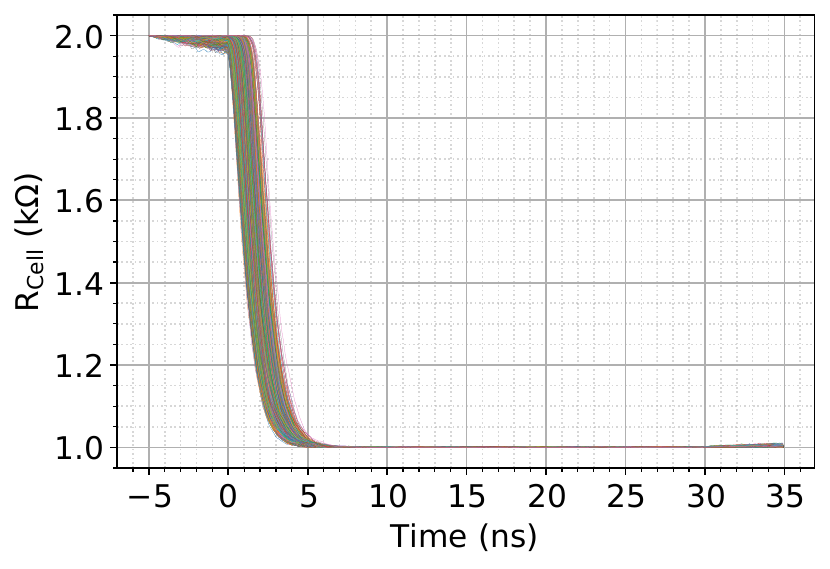}
        \caption{}
        \label{fig:logic1_resistance}
    \end{subfigure}%
    \vspace{\medskipamount}
    \begin{subfigure}[t]{1\textwidth}
        \centering
        \includegraphics[trim=0.25cm 0.25cm 0.25cm 0.25cm, clip, width = 0.98\textwidth]{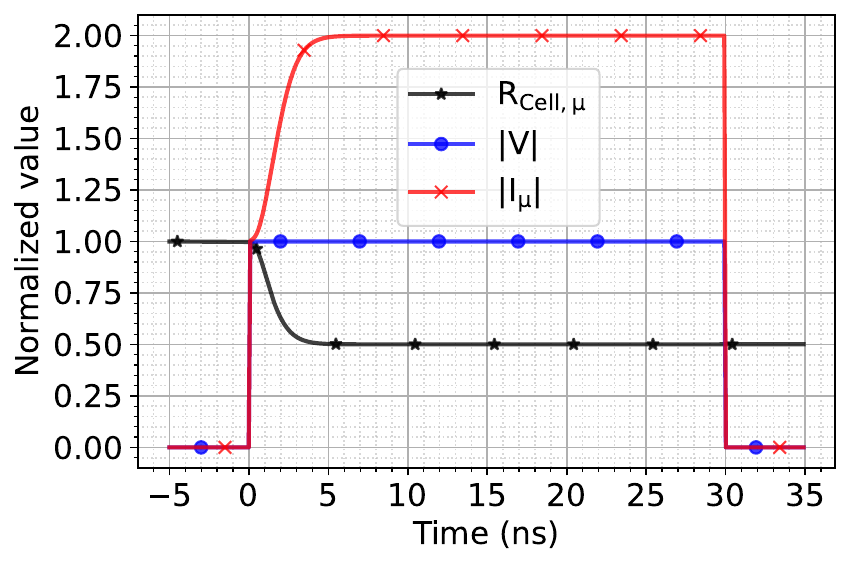}
        \caption{}
        \label{fig:logic1_resistance_voltage_current}
    \end{subfigure}%
    \vspace{\medskipamount}
    \begin{subfigure}[t]{1\textwidth}
        \centering
        \includegraphics[trim=0.25cm 0.25cm 0.25cm 0.25cm, clip, width = 0.98\textwidth]{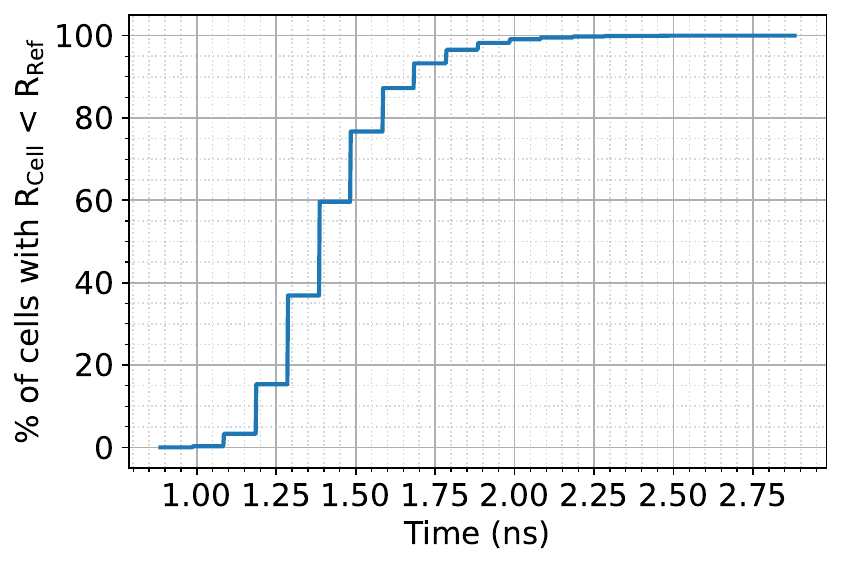}
        \caption{}
        \label{fig:logic1_cdf}
    \end{subfigure}
\end{minipage}
\caption{Switching characteristics obtained from $1M$ STT-MRAM cells over $t_{W}$, (a)--(c) low to high resistance switching and (d)--(f) high to low resistance switching. Fig. (a) and (d) present normalized cell resistance ($R_{Cell}$) distribution during switching. Fig. (b) and (e) show mean (normalized) cell resistance ($R_{Cell,\mu}$, dark curve), write voltage pulse ($|V|$, blue curve), and average write current ($I_{\mu}$, red curve) characteristics during switching. Fig. (c) and (f) present CDFs of $t_{W}$ at which cells switch their side (i.e., $R_{Low} \rightarrow R_{Ref} + \delta$ or  $R_{High} \rightarrow R_{Ref} - \delta$).}
\label{fig:Curr_Ana}
\vspace{-0.65cm}
\end{figure}
%\vspace{-0.1cm}

We summarize our simulation result in Fig. \ref{fig:Curr_Ana}, performed over $1M$ STT-MRAM cells. Figs. \ref{fig:logic0_resistance}--\ref{fig:logic0_cdf} represent MRAM switching characteristics for low to high resistance (normalized). On the other hand, \ref{fig:logic1_resistance}--\ref{fig:logic1_cdf} show the opposite switching characteristics (i.e., $R_{High} \rightarrow R_{Low}$). Fig. \ref{fig:logic0_resistance} and \ref{fig:logic1_resistance} show the cell resistance ($R_{Cell}$) over $t_{W}$ for all $1M$ memory cells. These figures prove that switching the low resistance state (logic ``1") to the high resistance state (logic ``0") is more critical (higher transition time) than the high to low resistance state switching, which we also observe from our experimental result. The $t_{W}$ should be sufficiently large to guarantee proper switching of all cells under all kinds of physical variation. From this simulation statistics, if we want to guarantee $99.99\%$ write success (${<}1$ write failure out of $10K$ write operations), the $t_{W}$ should be at least $15.45ns$. The $t_{W}$ also depends on operating conditions. For example, with $10\%$ voltage variation (\cite{synopsy:IRDrop}) and $99.99\%$ write accuracy, the $t_{W}$ increases to $17.45ns$. In this discussion, we will consider $t_{W}=17.45ns$ as the write time for standard operation.
% \textcolor{green}{17.45ns for what?} However, in practice, choosing $t_{W}$ is much more constrained ($<1$ write failure out of million write operation) with an additional margin.}

Figs. \ref{fig:logic0_resistance_voltage_current} and \ref{fig:logic1_resistance_voltage_current} represent mean (normalized) cell resistance ($R_{Cell,\mu}$), write pulse voltage ($|V|$), and write current ($I_{\mu} = \frac{|V|}{R_{Cell, \mu}}$). In these figures, mean write energy is proportional to the area under $I_{\mu}$ curve (as $|V|$ is constant). Fig. \ref{fig:logic0_cdf} and \ref{fig:logic1_cdf} represent the cumulative distribution function (CDF) of $t_{W}$ at which the memory cells switch the side of $R_{Ref}$ (i.e., $R_{Low} \rightarrow R_{Ref} + \delta$ or  $R_{High} \rightarrow R_{Ref} - \delta$). For $R_{High} \rightarrow R_{Low}$, all memory cells switch their side of $R_{Ref}$ within $t_{W} < 2.9ns$. However, for $R_{Low} \rightarrow R_{High}$, $99\%$ of cells switch their side of $R_{Ref}$ within $t_{W} < 10.2ns$ and $95\%$ of cells switch their side within $t_{W} < 8.45ns$. Therefore, in this simulation, applying $t_{W} = 8.45ns$ would produce almost similar to our experimental result (obtained at $t_{W} = 5ns$). Fig. \ref{fig:logic0_resistance_voltage_current} and \ref{fig:logic1_resistance_voltage_current} show that ($I_{\mu}$ curve), with $t_{W} = 8.45ns$, the mean energy saving would be ${\sim}40\%$ for $R_{Low} \rightarrow R_{High}$ transition and ${\sim}55\%$ for $R_{High} \rightarrow R_{Low}$ transition (considering $100\%$ energy at $t_{W}=17.45ns$). Therefore, the average energy saving would be ${\sim}47.5\%$, considering the uniform distribution of logic ``0" and logic ``1" in application data. Note that the energy-saving would be more for a tighter estimation of $t_{W}$. 

\begin{table}[ht!]
\caption{Comparison of MRAM-based AC frameworks.}
\setcellgapes{1pt}%parameter for the spacing
\captionsetup{justification=centering, margin= 0cm}
\vspace{-0.3cm}
\makegapedcells
\centering
\setlength\tabcolsep{10pt} 
\resizebox{0.48\textwidth}{!}
{
    \begin{tabular}{|l|c|c|c|}
    \hline
     & $\Delta t_{W}$ & $\Delta E$ & $BER$ \\ \hline
    Ranjan \textit{et al.} \cite{AC_STT_Roy_DATE, AC_STT_Roy} & -- & $35.9$ -- $57.8\%$ & $0.1$--$1\%$ \\ \hline
    Sayed \textit{et al.}$^{\mathrm{*}}$ \cite{AC_Tahoori} & $42.31\%$ & $42.5\%$ & $12\%$ \\ \hline
    This work$^{\mathrm{*}}$ & $53.80\%$ & $49.5\%$ & $5\%$ \\ \hline
    \multicolumn{4}{l}{$^{\mathrm{*}}$Considering minimum write accuracy at nominal $t_{W}$: ${\sim}99.995\%$} \\
    \end{tabular}
}
\label{Tab:Comp}
\vspace{-0.4cm}
\end{table}

Table \ref{Tab:Comp} compares our proposed technique with MRAM-based prior approximate computing frameworks. As previous works are based on simulating STT-MRAM, we use our STT-MRAM-based simulation result for a fair comparison. The \nth{2} (${\Delta}t_{W}$) and \nth{3} (${\Delta}E$) columns of the table represent the reduction of \textit{write pulse width} ($t_{W}$) and write energy. The \nth{4} column shows the error (bit error rate or $BER$) induced in approximate computing. The result shows that our proposed technique produces superior results over \cite{AC_Tahoori}, considering ${\Delta}t_{W}$, ${\Delta}E$, and $BER$. On the other hand, the AC framework proposed by Ranjan \textit{et al.} can save more power with smaller $BER$ \cite{AC_STT_Roy_DATE, AC_STT_Roy}; however, it does not improve the write speed. Additionally, our proposed technique can be readily used in COTS available MRAM and does not require any modification on system architecture.

\subsection{Summary of Results}
%------------------------------------------------------------------------------------------%

Overall, we draw the following main conclusions from the results.
%\vspace{-0.2cm}
\begin{enumerate} [leftmargin=*, topsep=0pt,itemsep=-1ex,partopsep=1ex,parsep=1ex]
    \item {The cell characterization is performed once in a lifetime before deploying the MRAM in the computing system. However, the appropriate selection is required for the training data and frequency to perform characterization.}
    \item {The output quality degradation occurs when approximate application data are placed at addresses containing both erroneous and error-free (without address selection). 
    %\st{due to the bit errors present in erroneous addresses in the} 
    However, this is the worst possible scenario when the entire memory is initialized with all 1's before writing. However, with the proposed address selection strategy, we can achieve almost $100\%$ of output quality.}
    \item {Contrarily, the same applications retain almost $100\%$ of quality even when put into addresses containing both erroneous and error-free in the case when the entire memory initialized with all 0's before writing.}
    \item {Commercial-grade MRAM is more vulnerable to reduced $t_W$ than industrial-grade MRAM %\st{due to internal MRAM architecture} as Commercial-grade MRAMs are more susceptible to noise.
    }
    \item {Errors that occurred at lower temperatures are a subset of the errors at high temperatures.}
    \item {Our simulation result shows that our proposed framework will also be applicable for STT-MRAM. The simulation also shows that the reduced \textit{write pulse width} ($t_W$) can also save on average ${\sim}47.5\%$ of write energy (considering $99.99\%$ write accuracy with nominal $t_{W}$) with virtually no loss in output quality for error-resilient applications.}
\end{enumerate}

%\vspace{-0.2cm}

%Hence, we can conclude from these results that our proposed AC framework is a much more fine-grained and accurate allocation strategy.

\vspace{-0.3cm}
%------------------------------------------------------------------------------------------%
\section{Discussions} \label{sec:discuss}
%------------------------------------------------------------------------------------------%

%------------------------------------------------------------------------------------------%
\subsection{Storage and Performance Overheads}
%------------------------------------------------------------------------------------------%

Our proposed framework requires allocating application data (critical and approximate) to first error-free addresses and moving forward to erroneous addresses based on the required application accuracy. We use a similar approach proposed in \cite{AC_dram}; however, our approach is simpler as the access policy of DRAM and MRAM is entirely different, and we do not require quality-wise erroneous address sorting. Note that the following strategy needs to be performed only for high-accuracy applications when no more accurate addresses are available. For this purpose, a custom memory allocator (resource manager) is required to track the erroneous addresses obtained from the characterization step. Next, based on the application requirements, it will allocate the user annotated critical and approximate data into virtual addresses using a critical bit for each address. Operating systems (OS) usually accomplish mapping virtual to physical addresses through a page table (with the assist of MMU). Hence, in our proposed scenario, the OS has to perform this additional responsibility of assigning virtual to physical addresses belonging to error-free (accurate) and erroneous addresses depending on the application requirement.

%\vspace{-0.13cm}
The OS and the MMU require additional logic to implement the proposed addresses selection strategy.
%However, a software-level enhancement is sufficient to perform the required task.
Towards this end, first, the OS can track the erroneous addresses using a custom data structure (consisting of critical bits).
%Second, an additional field in the MMU's page table can help to specify the erroneous addresses of each virtual address obtained from the custom allocator.
Then, at the start-up phase, the OS can utilize the custom data structure to modify the core map and subsequently utilize this modified mapping function to translate virtual to a physical address. 
%
%Then the OS can use the mentioned custom data structure to find a suitable physical frame for mapping each virtual address.  Note that the custom data structure is used in conjunction with the core map.
Note that the modified mapping is used in conjunction with the original core map. 
%Obtaining the physical address from the custom data structure, the OS checks the core map to determine the availability of the selected address. The decision to allocate to the currently selected physical address or skip it for the next one is made after that. Thus, b
Before performing virtual to physical address mapping, first, the OS will check whether the virtual address is critical or not for the required application. If it is critical, the OS will check for the error-free address based on the application requirement as specified by the system designer. In addition to that, one can use additional flag bits on the cache memory (similar to the \textit{valid} bit) to identify if a data block is erroneous or accurate. This flag bit can be directly copied from the custom data structure in parallel to the main memory operations. The OS can utilize this flag bit during the memory allocation for further optimization.
%Note that the performance penalty is insignificant for looking up the custom data structure compared to the baseline implementation. However, simultaneously looking up to multiple table entries to determine the correct physical address can reduce this mentioned latency. 
It is expected that the OS will completely automate the entire allocation process for the proposed framework. %After completion of application execution, it is required to free all the associated physical addresses.
Hence, hardware and OS support are essential for adapting approximate MRAM to reality in sophisticated high-end embedded computing systems. However, a software-based memory management can further simplify our framework as it does not require any virtual to physical address mapping \cite{zagieboylo2020cost}. Note that for a 1GByte main memory (with 32Byte block size), the size of the custom data structure would be 4MByte.
%\textcolor{blue}{To keep a record of the erroneous addresses, we require an additional column in the bitmap where this extra column corresponds to the erroneous addresses. However, considering a larger page size can improve this storage efficiency. The virtual to physical memory address translation is performed by keeping the bitmap in persistent storage and referring by the OS. Moreover, the performance overhead depends entirely on the software and hardware architecture and the application characteristics. Furthermore, the overhead due to reading and processing the custom data structure is insignificant comparing with the overall application execution time, energy, and performance point of view.
It is worth mentioning that the latency and energy overheads for the MRAM characterization are negligible as the characterization needs to be performed only once before deploying the MRAM in the computing system.

\vspace{-0.3cm} 
%------------------------------------------------------------------------------------------%
\subsection{Critical Data Protection: MRAM as a Cache}
%------------------------------------------------------------------------------------------%

A significant challenge of approximate storage usage is: most of the highly amenable approximate computing paradigm applications also have a mixture of control data (i.e., critical data) that is intolerable to any errors. Hence imposing approximation on these data makes them unreliable. For example, the instruction cache requires to be entirely error-free. Although various solutions are proposed to protect the critical data part \cite{AC_Tahoori}, the critical data size is minimal compared with the non-critical counterpart. Therefore, designing a heterogeneous data cache memory array with different static (design-time) and dynamic (runtime) configurations to make critical data error-resilient guarantees error-free operation. However, this requires (i) fabrication parameters modification (ii) complex cache controller for proper data allocation in different arrays. Contrarily, a more straightforward solution for critical data protection is either using multiple copies of this data content or an error correction code (ECC), where data protection is performed through additional check bit(s). However, the overhead due to the additional bit-cells for the mentioned approaches is minimal for these significantly smaller critical data sizes \cite{AC_Tahoori}. Besides, considerably lower write accesses to the instruction cache help the critical data protection considerably more manageable for our proposed framework.

\vspace{-0.3cm}
%------------------------------------------------------------------------------------------%
\subsection{Applicability to Next Generation MRAMs}
%------------------------------------------------------------------------------------------%

We used COTS toggle MRAM, which is designed to act like SRAM. On the other hand, STT-MRAM performs like a persistent DRAM. %\st{The main advantage of STT-MRAM over Toggle MRAM is the ability to scale, hence achieving higher densities at a lower price, providing a significant reduction in switching energy and consuming low power compared to Toggle MRAM devices. In STT-MRAM, the current is injected perpendicularly into the magnetic tunnel junction (MTJ), and the read/write operation is performed through the same path \mbox{\cite{SOT}}.} 
The \nth{1} generation STT-MRAM structure uses an in-plane MTJ (iMTJ), whereas \nth{2} generation STT-MRAM devices use a more optimized structure known as perpendicular MTJ (pMTJ) in which the magnetic moments are perpendicular to the silicon substrate surface \cite{khvalkovskiy2013basic}. Therefore, \nth{2} generation STT-MRAM is more scalable and cost-competitive than \nth{1} generation STT-MRAM, thus a more promising technology to replace DRAM and other memory technologies. Our simulation result presented in Sect. \ref{subsubsec:powImpv} (based on \nth{2 } generation STT-MRAM) confirms that our proposed AC framework is also applicable for STT-MRAM. Additionally, the DRAM-like operation of STT-MRAM can provide further opportunities to improve performance. The DRAM read/write operations are generally performed in burst mode, where 4 or 8 data words are accessed (read/write) in a single command. In burst mode, only the first write operation requires full write time, whereas the next consecutive write operations only require an additional write pulse for each data word. Therefore, reducing write pulse width ($t_{W}$) can improve huge performance.

On top of that, unlike STT-MRAM, the new spin-orbit torque MRAM (SOT-MRAM) technology switched the free magnetic layer by injecting an in-plane current in an adjacent SOT layer (typically made of heavy metal). Due to the current injection geometry, the read and write paths are decoupled. SOT-MRAM offers better performance in terms of speed (sub-ns switching speed), endurance, power,  and read stability at the cost of a slightly degraded density. %\st{The newest MRAM technologies only focus on device architecture level improvement to ensure improved density, endurance, or cost;} 
As the SOT-MRAM has a similar switching characteristic as the previous generation of MRAMs \cite{SOT}, we believe our system-level proposed approach also applies to future MRAM generations. In the future, we plan to explore the SOT-MRAM experimentally (or with simulation) to confirm our hypothesis.

\vspace{-0.3cm}
%------------------------------------------------------------------------------------------%
\section{Conclusion} \label{sec:end}
%------------------------------------------------------------------------------------------%

This paper proposes an efficient approximate computing framework to evaluate the applicability of non-volatile COTS MRAM, a promising candidate for future computing platforms, through memory address's error characterization by utilizing the internal write latency variation of MRAM to improve the power efficiency. Our methodology provides an optimal system-level implementation yielding a favorable performance and power vs. quality trade-off for error-resilient applications by devising efficient segregation strategies for the erroneous addresses while allocating approximate and critical data systematically. 
% during page mapping 
Our experimental results reveal that MRAM-based memory systems can achieve substantial power and performance benefits for negligible or no loss in application output quality. Furthermore, other emerging memory technologies can also adopt the proposed address selection strategy, a promising framework aspect.

\vspace{-0.4cm}

% or


% \appendices

% % you can choose not to have a title for an appendix
% % if you want by leaving the argument blank
% \section{}

% % use section* for acknowledgment
% \ifCLASSOPTIONcompsoc
%   % The Computer Society usually uses the plural form
%   \section*{Acknowledgments}
% \else
%   % regular IEEE prefers the singular form
% \fi


\ifCLASSOPTIONcaptionsoff
  \newpage
\fi



%\bibliographystyle{IEEEtran}

% \end{thebibliography}
\bibliographystyle{IEEEtran}
\bibliography{IEEEabrv,ref}

\vspace{-1.2cm}
\vskip 0pt plus -1fil 
\begin{IEEEbiography}
[{\includegraphics[width=1in,height=1.25in,clip, trim=0cm 0cm 0cm 0cm, keepaspectratio]{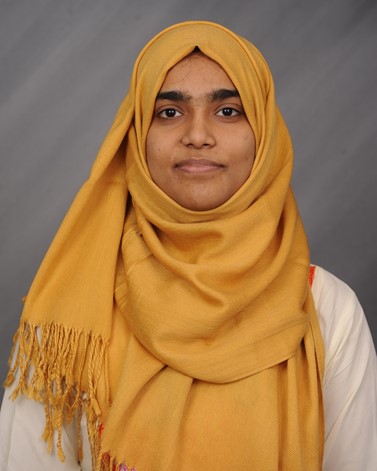}}]
{Farah Ferdaus} (S'20) received her M.S. degree in Electrical and Computer Engineering from the University of New Hampshire in 2018. She is currently working towards her Ph.D. in Electrical and Computer Engineering at Florida International University. Her research interests include performance enhancement and security solutions of emerging memories; privacy and security issues of existing memories; wireless communications and networks.
\end{IEEEbiography}
\vspace{-1.2cm}
\vskip 0pt plus -1fil 
\begin{IEEEbiography}
[{\includegraphics[width=1in,height=1.25in,clip,keepaspectratio]{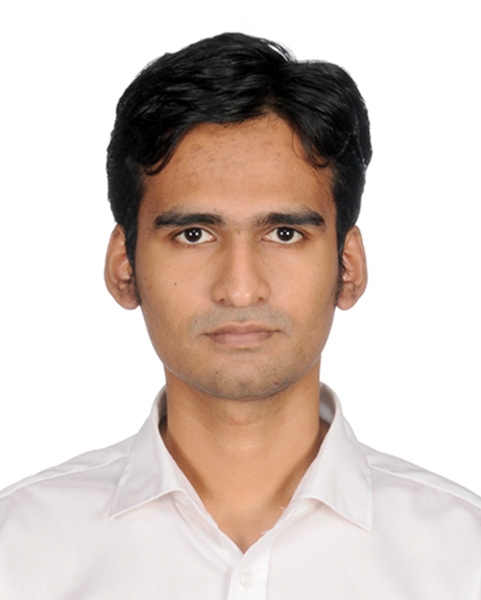}}]
{B. M. S. Bahar Talukder} (S'18) received his Ph.D. degree in Electrical and Computer Engineering at Florida International University. %He received his Bachelor's degree from Bangladesh University of Engineering and Technology, Dhaka, Bangladesh. 
His primary research interests include hardware security, secured computer architecture, machine-learning application in system security, and emerging memory technologies.
\end{IEEEbiography}
\vspace{-1.2cm}
\vskip 0pt plus -1fil 
\begin{IEEEbiography}
[{\includegraphics[width=1in,height=1.25in,clip, trim=0cm 0cm 0cm 0cm, keepaspectratio]{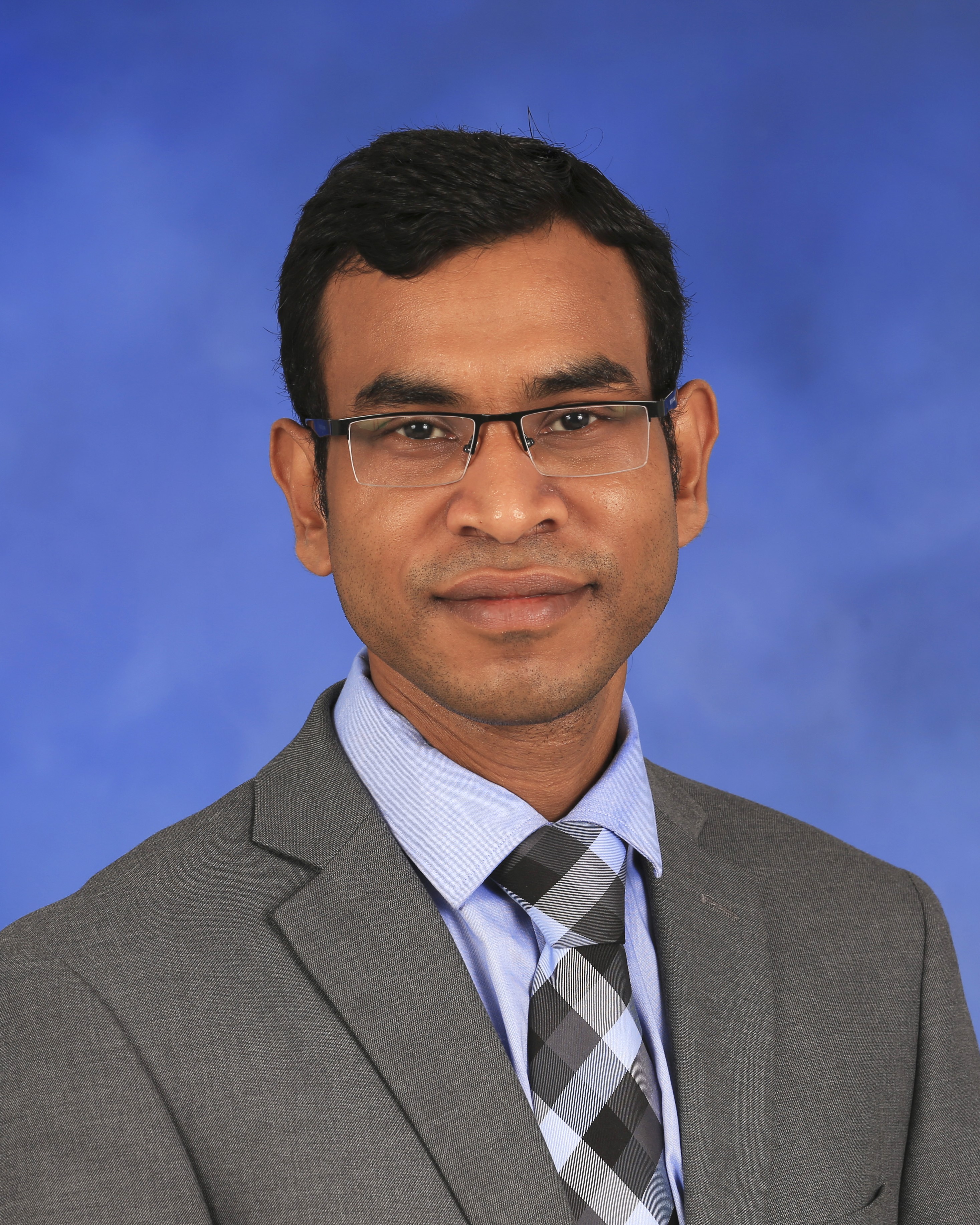}}]
{Md Tauhidur Rahman} (S'12–M'18) is an Assistant Professor in the Department of Electrical and Computer Engineering at Florida International University (FIU). He received his Ph.D. degree in Computer Engineering from the University of Florida in 2017. His current research interests include hardware security and trust, memory system, machine learning applications, embedded security, and reliability.
\end{IEEEbiography}

%\newpage



%\vfill


\end{document}